\documentclass[journal]{IEEEtran}
\usepackage{amsmath,amsfonts}
\usepackage{algorithmic}
\usepackage{algorithm}
\usepackage{array}
\usepackage{textcomp}
\usepackage{stfloats}
\usepackage{url}
\usepackage{verbatim}
\usepackage{graphicx}
\usepackage{cite}
\usepackage{amsthm}
\usepackage{amssymb}
\usepackage{epsfig}
\usepackage{graphicx}
\usepackage{balance}
\usepackage{epstopdf}
\usepackage{multicol}
\usepackage{multirow}
\usepackage{threeparttable}
\usepackage{booktabs}
\usepackage{cite}
\usepackage{cases}
\usepackage{cuted}
\usepackage{subfig}

\usepackage{amsmath}
\allowdisplaybreaks[2]

\usepackage{microtype}
\usepackage[%
left=1.2cm,
right=1.2cm,
top=1.3cm,
bottom=1.3cm
]{geometry}

\begin{document}
	
\setlength{\abovedisplayskip}{4pt plus 1pt minus 1pt}
\setlength{\belowdisplayskip}{4pt plus 1pt minus 1pt}
\setlength{\abovedisplayshortskip}{2pt plus 1pt minus 1pt}
\setlength{\belowdisplayshortskip}{2pt plus 1pt minus 1pt}
\setlength{\textfloatsep}{4pt} 
\setlength{\floatsep}{2pt}     
\setlength{\intextsep}{4pt}    
	
\title{Node-Oriented Proactive Spectral Modulation: A Unified Fractional Framework for Graph Signal Denoising}

\author{Manjun~Cui, Zhichao~Zhang,~\IEEEmembership{Member,~IEEE}, and Yangfan~He
	\thanks{This work was supported in part by the Open Foundation of Hubei Key Laboratory of Applied Mathematics (Hubei University) under Grant HBAM202404; in part by the Foundation of Key Laboratory of System Control and Information Processing, Ministry of Education under Grant Scip20240121; in part by the Startup Foundation for Introducing Talent of Nanjing Institute of Technology under Grant YKJ202214; and in part by the Open Foundation of Jiangsu Province Engineering Research Center of IntelliSense Technology and System under Grant ITS202502. \emph{(Corresponding author: Zhichao~Zhang.)}}
	\thanks{Manjun~Cui is with the School of Mathematics and Statistics, Nanjing University of Information Science and Technology, Nanjing 210044, China (e-mail: cmj1109@163.com).}
	\thanks{Zhichao~Zhang is with the School of Mathematics and Statistics, Nanjing University of Information Science and Technology, Nanjing 210044, China, with the Hubei Key Laboratory of Applied Mathematics, Hubei University, Wuhan 430062, China, and also with the Key Laboratory of System Control and Information Processing, Ministry of Education, Shanghai Jiao Tong University, Shanghai 200240, China (e-mail: zzc910731@163.com).}
	\thanks{Yangfan~He is with the School of Communication and Artificial Intelligence, School of Integrated Circuits, Nanjing Institute of Technology, Nanjing 211167, China, and also with the Jiangsu Province Engineering Research Center of IntelliSense Technology and System, Nanjing 211167, China (e-mail: Yangfan.He@njit.edu.cn).}}

\markboth{IEEE TRANSACTIONS ON SIGNAL PROCESSING}
{Shell \MakeLowercase{\textit{et al.}}: Bare Demo of IEEEtran.cls for Journals}


\maketitle

\begin{abstract}
	Graph signal denoising is a fundamental task in graph signal processing. While the node-oriented filtering approach enhances spatial adaptability, it suffers from spectral rigidity due to its reliance on the graph Fourier transform. Conversely, emerging fractional-domain transforms provide crucial spectral flexibility but are fundamentally limited by their globally shared filtering paradigm, failing to accommodate localized topological variations. To bridge this gap, this paper proposes a generalized node-oriented fractional filtering (NOFF) framework that seamlessly integrates localized spatial adaptability with proactive spectral modulation across various fractional transforms. However, straightforwardly assigning independent full-rank filters to all vertices incurs a prohibitive parameter space, leading to severe overfitting on random noise. To mitigate this, we introduce the low-rank NOFF (LRNOFF) architecture. By imposing a strict low-rank constraint, LRNOFF inherently acts as a powerful implicit regularizer, preventing noise memorization and ensuring the extraction of robust spectral bases. Furthermore, we develop an efficient computational implementation termed LRNOFF-Fast, which drastically reduces computational and memory overhead while preserving theoretical optimality. Experiments on real-world datasets demonstrate that the proposed framework achieves state-of-the-art performance.
\end{abstract}

\begin{IEEEkeywords}
	Fractional-domain transforms, graph signal denoising, low-rank approximation, node-oriented filtering. 
\end{IEEEkeywords}

\section{Introduction}
\indent In recent years, data generated from diverse real-world applications, such as sensor networks, transportation systems, social media, and biological networks, are increasingly represented on irregular, non-Euclidean domains. To effectively analyze such complex data, graph signal processing (GSP) has emerged as a powerful and unified mathematical framework by extending classical SP concepts to graph-structured data \cite{Ortega08,ortega22introduction}. GSP has demonstrated remarkable efficacy in a multitude of fundamental tasks, including signal reconstruction, sampling, semi-supervised classification, and prediction \cite{Wang24,Li23,Song23 ,Bao22, DONG2019106972, Saboksayr21}. Among these tasks, graph signal denoising plays a foundational and indispensable role \cite{AUnified21Ma,Onuki16Graph,Chen21Graph,Yagan20Spectral}. Since real-world graph signals collected from physical environments are inevitably corrupted by measurement errors, sensing anomalies, or transmission noise, effective denoising serves as a critical preprocessing step that fundamentally dictates the performance and reliability of downstream analysis and graph representation learning.

\indent In GSP, the graph Fourier transform (GFT) provides the fundamental spectral tools for graph signal denoising \cite{Sandryhaila13,Domingos20,Qi22,Yagan16Aspectral}. By mapping signals into the graph spectral domain, denoising can be achieved by designing specific spectral responses, such as low-pass or band-pass filters. A fundamental premise of these conventional approaches is that the filters are globally shared, meaning an identical spectral response is rigidly applied to all vertices across the entire graph. However, real-world graph data typically exhibit spatial irregularity, structural heterogeneity, and localized varying patterns. Consequently, a globally shared filter suffers from a one-size-fits-all limitation. It is fundamentally incapable of accommodating the highly varying local neighborhood topologies, thereby severely restricting its representational capacity and denoising efficacy in regions with complex structural characteristics.

\indent To overcome the limitations of global filtering, recent advances have introduced the concept of node-oriented graph filtering, which assigns customized spectral responses to individual vertices \cite{ZhengNode24}. By providing independent and flexible spectral weights for different nodes, this paradigm significantly enhances spatial adaptability, enabling a more accurate modeling of localized topological variations and node-specific characteristics. Consequently, it has demonstrated remarkable empirical success in complex graph representation learning tasks, particularly in semi-supervised node classification on graphs with severe heterophily \cite{ZhengNode24}. However, despite this spatial flexibility, the node-oriented method is predominantly established upon the standard GFT domain, thereby restricting the spectral representational capacity to a rigidly fixed set of graph spectral bases. When the underlying true signals and complex noise patterns are severely aliased within this standard spectral domain, merely allocating localized weights over fixed basis vectors is fundamentally insufficient.  The GFT lacks the essential degrees of freedom required to proactively decouple these entangled frequency components, ultimately limiting the denoising potential of current node-oriented methods in heavily corrupted scenarios.

\indent To overcome the spectral rigidity of the standard GFT, researchers have recently explored various fractional-domain graph transforms, such as the graph fractional Fourier transform (GFRFT) \cite{Chen26Hyper,Wang18,Ozturk21,gan2025windowed,Alik24Wiener}, the multiple-parameter GFRFT (MPGFRFT) \cite{cui2025multiple}, and the graph linear canonical transform (GLCT) \cite{li2025multi,Chen24garph,Zhang23Discrete}. These transforms break the fixed-basis limitation by introducing crucial extra degrees of freedom through tunable fractional parameters, enabling proactive spectral modulation to separate entangled signal and noise components that are otherwise inseparable in the standard GFT domain. By mapping graph signals into optimal intermediate transform spaces, fractional-domain filtering has demonstrated remarkable empirical success across a wide range of complex scenarios, including sensor network denoising, image reconstruction, and biological data analysis \cite{YAN2022Multi,yan2021windowed,alikacsifouglu2025joint,Alikasifoglu24Unified,Tseng24TV,Tseng24AFractional,Li23Review}. However, despite their spectral superiority, most existing fractional-domain filters are strictly designed as globally shared fractional filters, meaning they rigidly apply the same transform orders and spectral coefficients to every vertex across the graph. This global-sharing paradigm prevents these transforms from effectively accommodating the spatial irregularity and structural heterogeneity of complex networks. Such a fundamental limitation naturally raises a compelling question: how can we systematically integrate fractional transforms with node-oriented mechanisms to achieve both spatial adaptability and spectral flexibility?

\begin{figure}[htbp]
	\centering
	\includegraphics[width=0.9\linewidth]{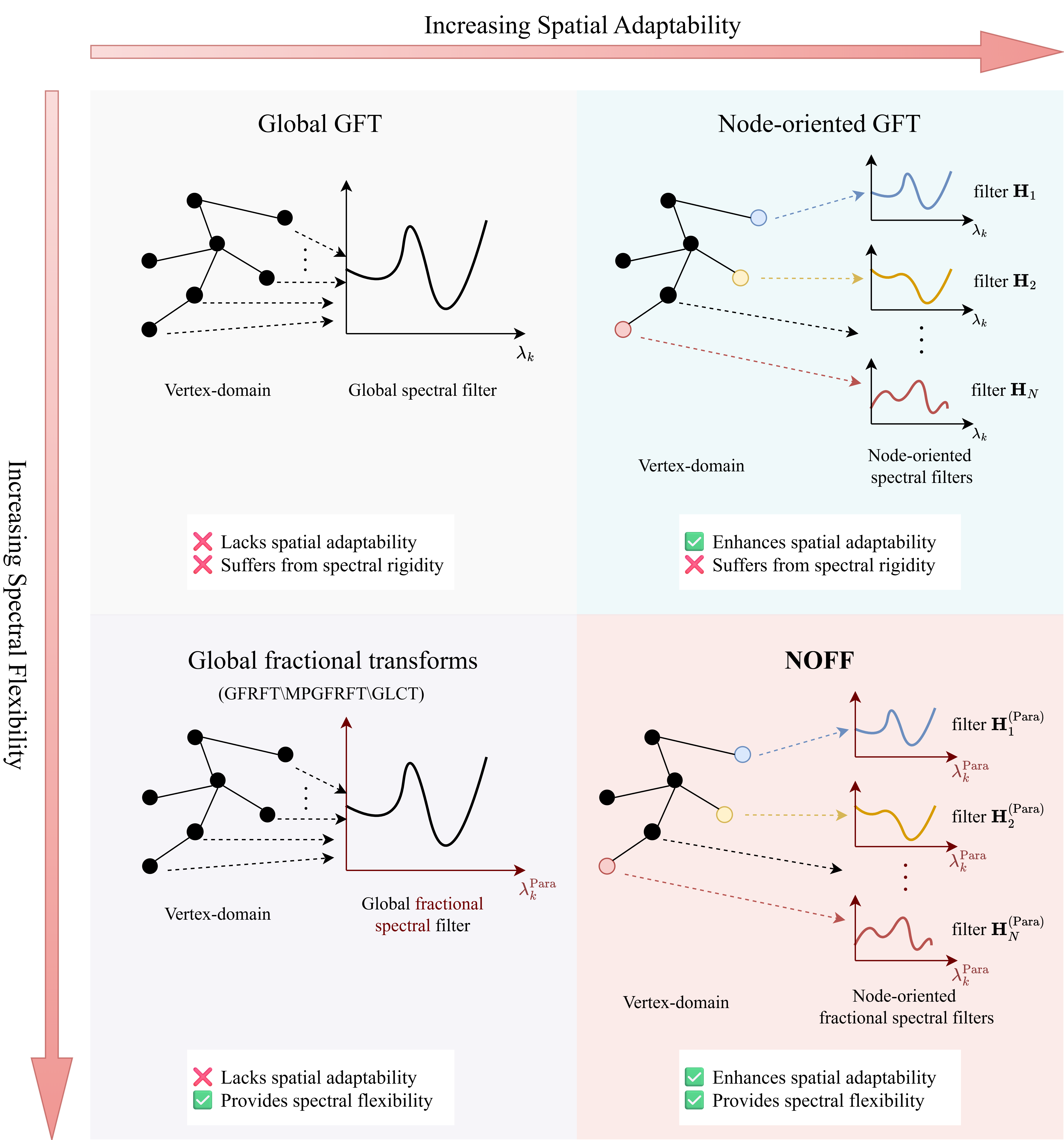} 
	\caption{Conceptual comparison of different graph spectral filtering paradigms.}
	\label{fig:motivation}
\end{figure}

\indent Motivated by this fundamental question, this paper proposes a unified node-oriented fractional filtering (NOFF) framework, which systematically integrates the localized spatial adaptability of node-oriented filtering with the superior spectral flexibility of fractional-domain transforms. As conceptually illustrated in Fig. \ref{fig:motivation}, the proposed NOFF architecture elegantly overcomes the dual limitations of existing methods, establishing a comprehensive paradigm that achieves both properties simultaneously. By generalizing node-wise spectral modulation from the rigid GFT domain to a wide range of fractional domains, including GFRFT, MPGFRFT, and GLCT, the NOFF framework allows each vertex to adaptively learn its own personalized filtering response within an optimal intermediate transform space. However, a straightforward implementation of the NOFF framework on large-scale graphs encounters a significant hurdle. Assigning unconstrained filters to all $N$ vertices results in a prohibitive  parameter space of $N^2$. This immense parameterization not only leads to catastrophic memory consumption but also causes the model to memorize random graph noise rather than learning robust, physically meaningful filtering patterns, thereby severely compromising its generalization capability.

\indent To mitigate these severe challenges on large-scale graphs with massive vertices, this paper introduces a low-rank node-oriented fractional filtering (LRNOFF) architecture. Rather than learning unconstrained dense filters, LRNOFF imposes a structural low-rank bottleneck on the ideal localized filter matrix. Crucially, this structural constraint inherently acts as an implicit regularizer during the end-to-end training process. By deliberately restricting the hypothesis space, it effectively prevents the model from memorizing high-frequency random noise, forcing the network to capture only the most essential and robust spectral filtering patterns. Ultimately, this synergistic design establishes a unified fractional filtering framework that seamlessly bridges the theoretical optimality of node-oriented spectral modulation with the robust regularization required for complex graph signal processing. The main contributions of this paper are summarized as follows:
\begin{itemize}
	
	\item \textbf{NOFF framework:} We propose the NOFF framework, which introduces localized spatial adaptability into fractional-domain transforms (GFRFT, MPGFRFTs and GLCTs). This fundamentally overcomes both the spectral rigidity of standard node-oriented methods and the spatial limitations of globally shared fractional filters.
	
	\item \textbf{LRNOFF framework and implicit regularization:} We introduce the LRNOFF architecture by factorizing the ideal unconstrained filter matrix into a low-rank structure. We illustrate that this constraint naturally serves as an implicit regularizer, which drastically reduces the parameter space and prevents noise memorization.
	
	\item \textbf{Scalable LRNOFF-Fast implementation:} To facilitate practical deployment on massive real-world networks, we develop a scalable computational approach termed LRNOFF-Fast. 
	
	\item \textbf{State-of-the-art denoising performance:} Experiments on real-world datasets demonstrate that our proposed framework consistently achieves state-of-the-art performance. The results explicitly validate the superiority of node-oriented mechanisms over globally shared filtering, and fractional-domain transforms over the standard GFT, while significantly outperforming advanced graph neural network (GNN) baselines.
	
\end{itemize}

\begin{figure*}[htbp]
	\centering
	\includegraphics[width=0.8\textwidth]{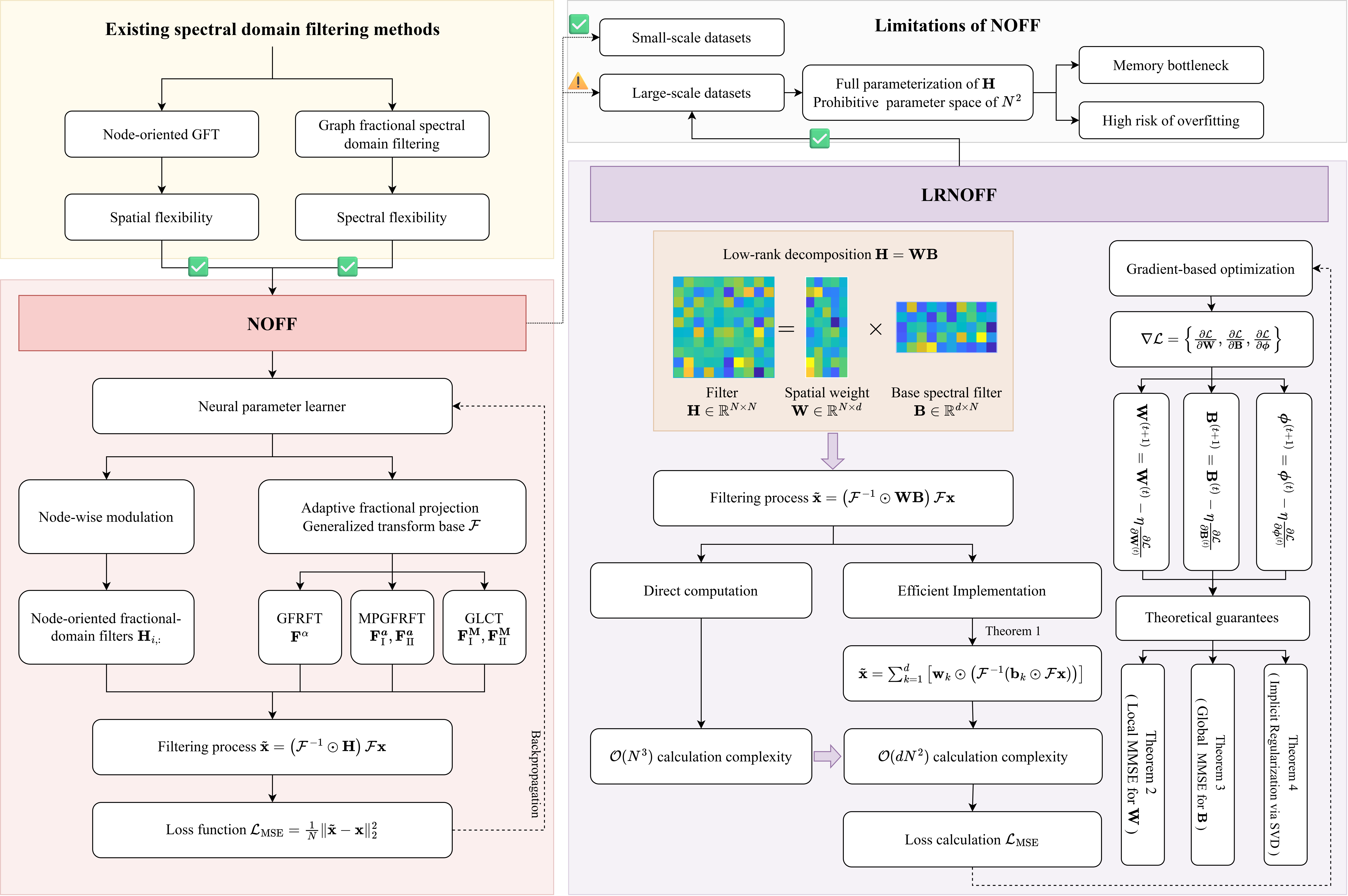}
	\caption{The overall technical architecture of the proposed NOFF and its scalable LFNOFF.}
	\label{fig:overall_architecture}
\end{figure*}

\indent  The remainder of this paper is organized as follows. Section \ref{sec2} introduces the essential preliminary concepts and notations. Section \ref{sec3} presents the theoretical formulation of the proposed NOFF framework. Section \ref{sec4} introduces the scalable LRNOFF architecture, detailing its mathematical formulation and efficient computational strategy. Section \ref{sec5} provides a comprehensive numerical evaluation of the proposed frameworks on both small-scale and large-scale graph datasets. Finally, Section \ref{sec6} concludes this paper. The comprehensive flowchart of the proposed NOFF and LRNOFF architectures is illustrated in Fig. \ref{fig:overall_architecture}. All the technical proofs of our theoretical results are relegated to the Appendix parts.

\section{Preliminaries} \label{sec2}
\indent In this section, we briefly introduce some basic concepts of GSP and several fractional-domain transforms.

\subsection{Graph signals and GFT}
\indent Consider an undirected weighted graph $\mathcal{G} = (\mathcal{V}, \mathcal{E}, \mathbf{W})$, where $\mathcal{V}$ is the set of $N$ vertices, $\mathcal{E}$ is the set of edges, and $\mathbf{W} \in \mathbb{R}^{N \times N}$ is the weighted adjacency matrix. The degree matrix is $\mathbf{D} = \mathrm{diag}(d_1, d_2, \cdots, d_{N})$ with $d_i = \sum_{j=1}^N W_{ij}$, and the Laplacian matrix is given by $\mathbf{L} = \mathbf{D} - \mathbf{W}$. A graph signal is defined as a vector $\mathbf{x} = [x_1, x_2, \cdots, x_{N}]^{\mathrm{T}} \in \mathbb{R}^N$.

\indent Let $\mathbf{S}$ denote a graph shift operator (GSO) (e.g., the adjacency matrix $\mathbf{W}$ or the Laplacian matrix $\mathbf{L}$). In this paper, we assume that $\mathbf{S}$ is a real symmetric matrix, which ensures it can be orthogonally diagonalized as $\mathbf{S} = \mathbf{V} \mathbf{\Lambda} \mathbf{V}^{\mathrm{H}}$, where $\mathbf{V}$ consists of the eigenvectors and $\mathbf{\Lambda}=\mathrm{diag}\left(\lambda_0, \lambda_1, \cdots, \lambda_{N-1}\right)$ is the diagonal matrix of eigenvalues. The GFT of signal $\mathbf{x}$ is defined as  
\begin{align}
	\hat{\mathbf{x}} = \mathbf{F} \mathbf{x},
\end{align}
where $\mathbf{F}=\mathbf{V}^{\mathrm{H}}$ is the GFT matrix.

\subsection{Fractional-domain graph transforms}
\indent To break through the limitation of the standard GFT, various fractional-domain transforms have been developed by generalizing the graph spectral domain.  

\indent By introducing a fractional order parameter $\alpha \in \mathbb{R}$ to provide a flexible intermediate domain between the vertex and spectral representations, the GFRFT is proposed.

\indent \emph{Definition 1 (GFRFT):} Given a fractional order parameter $\alpha \in \mathbb{R}$, the GFRFT matrix $\mathbf{F}^{\alpha}$ is defined as the fractional power of the standard GFT matrix \cite{Wang17}
\begin{equation}
	\mathbf{F}^{\alpha} = \mathbf{V} \mathbf{\Lambda}^{\alpha} \mathbf{V}^{\mathrm{H}},
\end{equation}
where $\mathbf{\Lambda}^{\alpha} = \mathrm{diag}\left(\lambda_0^{\alpha}, \lambda_1^{\alpha}, \cdots, \lambda_{N-1}^{\alpha}\right)$. For a graph signal $\mathbf{x}$, its $\alpha$-th order GFRFT is given by
\begin{align}
	\widehat{\mathbf{x}}_{\alpha} = \mathbf{F}^{\alpha} \mathbf{x}.
\end{align}

\indent Furthermore, by assigning independent fractional orders to different spectral components, the MPGFRFT is proposed in our previous work \cite{cui2025multiple}. This multiple-parameter mechanism provides higher degrees of freedom, inherently enabling proactive and fine-grained spectral modulation for complex graph signals. The MPGFRFT is defined in two forms.

\indent \emph{Definition 2 (MPGFRFT-I):} Given an order vector $\boldsymbol{a}=(a_0,a_1,\cdots,a_{N-1})$, the MPGFRFT-I of signal $\mathbf{x}$ is defined as
\begin{align}
	\widehat{\mathbf{x}}_{\boldsymbol{a}}^{\rm I} = \mathbf{F}_{\rm I}^{\boldsymbol{a}} \mathbf{x},
\end{align}
where the MPGFRFT-I matrix $\mathbf{F}_{\rm I}^{\boldsymbol{a}}$ is given by
\begin{align} \label{MPGFRFT-I matrix1}
	\mathbf{F}_{\rm I}^{\boldsymbol{a}} = \mathbf{V} \boldsymbol{\Lambda}_F^{\boldsymbol{a}} \mathbf{V}^{-1},
\end{align}
with $\boldsymbol{\Lambda}_F^{\boldsymbol{a}}\triangleq \mathrm{diag}(\lambda_0^{a_0},\lambda_1^{a_1},\cdots,\lambda_{N-1}^{a_{N-1}})$.

\indent \emph{Definition 3 (MPGFRFT-II):} Given an order vector $\boldsymbol{a}=(a_0,a_1,\cdots,a_{N-1})$, the MPGFRFT-II of signal $\mathbf{x}$ with respect to the order vector $\boldsymbol{a}$ is defined as
\begin{align}
	\widehat{\mathbf{x}}_{\boldsymbol{a}}^{\rm II} = \mathbf{F}_{\rm II}^{\boldsymbol{a}} \mathbf{x},
\end{align}
where the MPGFRFT-II matrix $\mathbf{F}_{\rm II}^{\boldsymbol{a}}$ is expressed as 
\begin{align}  \label{MPGFRFT-II matrix}
	\mathbf{F}_{\rm II}^{\boldsymbol{a}} = \sum\limits_{n=0}^{N-1}C_{n,a_n}^{\rm II}\mathbf{F}^{n},
\end{align}
with the coefficients
\begin{align} \label{coefficients-II}
	C_{n, a_{n}}^{\mathrm{II}}=\sum\limits_{j=0}^{N-1} p_{n+1,j+1} \lambda_{j}^{a_{n}}.
\end{align}
The constant matrix $\mathbf{P} = (p_{i j})$ is 
\begin{align}  \label{eq:P}
	\mathbf{P}=\left(p_{i j}\right) \triangleq\left(\begin{array}{cccc}
		1 & \lambda_{0} & \cdots & \lambda_{0}^{N-1} \\
		1 & \lambda_{1} & \cdots & \lambda_{1}^{N-1} \\
		\vdots & \vdots & \ddots & \vdots \\
		1 & \lambda_{N-1} & \cdots & \lambda_{N-1}^{N-1}
	\end{array}\right)^{-1}.
\end{align}

\indent In addition to the aforementioned transforms, the GLCT introduces an alternative parameterized framework. For a given parameter matrix $\mathbf{M} = (a, b; c, d)$ satisfying $ad - bc = 1$, the continuous parameters are mapped as $\xi = \frac{ac+bd}{a^2+b^2}$, $\sigma = \sqrt{a^2+b^2}$, and $\alpha = \frac{2}{\pi}\cos^{-1}\left(\frac{a}{\sigma}\right) = \frac{2}{\pi}\sin^{-1}\left(\frac{b}{\sigma}\right)$. Based on centered discrete Hermite-Gaussian functions (CDDHFs), the GLCT is defined under different GSOs.

\indent \emph{Definition 4 (Lap-CDDHFs-GLCT):} Using the Laplacian matrix $\mathbf{L}$, the Lap-CDDHFs-GLCT of signal $\mathbf{x}$ is defined as
\begin{align}
	\widehat{\mathbf{x}}_{\mathbf{M}}^{\mathrm{I}} =\mathbf{F}_{\mathrm{I}}^{\mathbf{M}} \mathbf{x},
\end{align}
and the transformation matrix $\mathbf{F}_{\mathrm{I}}^{\mathbf{M}}$ is given by
\begin{align}
	\mathbf{F}_{\mathrm{I}}^{\mathbf{M}} = \mathcal{CM}_{\mathbf{L}}^{\xi} \mathbf{S}_{\mathbf{L}} \mathbf{P}_{\mathbf{L}} \mathbf{J}_{\mathbf{L}}^{\alpha} \mathbf{P}_{\mathbf{L}}^{-1},
\end{align}
where $\mathbf{P}_{\mathbf{L}}$ is the eigenvector matrix of $\mathbf{L}$, and $\mathbf{J}_{\mathbf{L}}$ is the diagonal eigenvalue matrix. The operator $\mathbf{S}_{\mathbf{L}}$ is the Laplacian-based graph scaling matrix, constructed as
\begin{align}
	\mathbf{S}_{\mathbf{L}} = \mathrm{diag}([\sigma^{-\varepsilon r_0}, \sigma^{-\varepsilon r_1}, \cdots, \sigma^{-\varepsilon r_{N-1}}]),
\end{align}
where $r_l$ ($l = 0, \cdots, N-1$) represents the eigenvalues corresponding to the eigenvector matrix $\mathbf{P}_{\mathbf{L}}$, and $\varepsilon$ represents the rate of scale change. The operator $\mathcal{CM}_{\mathbf{L}}^{\xi}$ represents the Laplacian-based graph chirp multiplication matrix
\begin{align}
	\mathcal{CM}_{\mathbf{L}}^{\xi} = \mathrm{diag}(\mathbf{P}_{\mathbf{L}} \hat{\mathbf{s}}_{\xi}),
\end{align}
where $\hat{\mathbf{s}}_{\xi}$ is the spectral chirp signal vector whose $k$-th element is given by $\hat{s}_{\xi}(k) = e^{-i\frac{\lambda_k^2}{\xi}}$.

\indent \emph{Definition 5 (wAdj-CDDHFs-GLCT):} Using the weighted adjacency matrix $\mathbf{W}$, the wAdj-CDDHFs-GLCT of signal $\mathbf{x}$ is defined as
\begin{align}
	\widehat{\mathbf{x}}_{\mathbf{M}}^{\mathrm{II}} = \mathbf{F}_{\mathrm{II}}^{\mathbf{M}} \mathbf{x},
\end{align}
and the transformation matrix $\mathbf{F}_{\mathbf{M}}^{\mathrm{II}}$ is given by
\begin{align}
\mathbf{F}_{\mathrm{II}}^{\mathbf{M}} = \mathbf{J}_{\mathbf{W}}^{\xi} \mathbf{P}_{\sigma} \mathbf{J}_{\mathbf{W}}^{\alpha} \mathbf{P}_{\mathbf{W}}^{-1},
\end{align}
where $\mathbf{P}_{\mathbf{W}}$ is the eigenvector matrix of $\mathbf{W}$, $\mathbf{J}_{\mathbf{W}}$ denotes the corresponding eigenvalue matrix, and $\mathbf{P}_{\sigma}$ is the eigenvector matrix of the scaled adjacency matrix $\frac{1}{\sigma}\mathbf{W}$.

\indent For simplicity of presentation in the subsequent sections, the term ``CDDHFs" will be omitted, and these two transforms will be abbreviated as Lap-GLCT and wAdj-GLCT, respectively.

\section{Node-Oriented Fractional-Domain Filtering} \label{sec3}
\indent In GSP, graph filters are typically designed as global operators shared across the entire graph. Given an observed signal $\mathbf{x}$, the global filtering process in the standard graph spectral domain can be expressed as
\begin{align}
	\mathbf{y} = \mathbf{V} \mathbf{H}_{\mathrm{diag}} \mathbf{V}^{\mathrm{H}} \mathbf{x},
\end{align}
where $\mathbf{H}_{\mathrm{diag}} = \mathrm{diag}\left(h_1, h_2, \cdots, h_{N}\right)$ is the filter. 

\indent However, the globally shared filter fails to adapt to the spatial heterophily of real-world graph data. To address this, Zheng \emph{et al.} \cite{ZhengNode24} proposed a node-oriented filtering mechanism that assigns independent spectral filtering weights to each vertex, significantly enhancing spatial adaptability. Although effective, this method is strictly confined to the standard graph spectral domain and lacks the capability for proactive spectral modulation. To simultaneously overcome spatial heterophily and spectral rigidity, we propose a node-oriented fractional-domain graph filtering method. By integrating spatial adaptation into the generalized fractional-domain systems defined in Section \ref{sec2}, our filter endows each vertex with exclusive spectral coefficients and leverages fractional degrees of freedom to perform proactive and fine-grained signal extraction.

\subsection{Node-oriented fractional filtering}
\indent For notational convenience, we uniformly use $\mathcal{F}$ and $\mathcal{F}^{-1}$ to denote any forward and inverse fractional-domain transform matrices, respectively. Depending on the specific generalized fractional-domain systems defined in Section \ref{sec2}, $\mathcal{F}$ can represent any of the five previously introduced matrix forms, i.e. $\{\mathbf{F}^{\alpha},\mathbf{F}_{\rm I}^{\boldsymbol{a}},\mathbf{F}_{\rm II}^{\boldsymbol{a}},\mathbf{F}_{\mathrm{I}}^{\mathbf{M}},\mathbf{F}_{\mathrm{II}}^{\mathbf{M}}\}$.

\indent In practical applications, the pure graph signal $\mathbf{x} \in \mathbb{R}^N$ is typically corrupted by noise. Let $\mathbf{y} \in \mathbb{R}^N$ denote the observed noisy signal, modeled as $\mathbf{y} = \mathbf{x} + \mathbf{n}$, where $\mathbf{n} \in \mathbb{R}^N$ represents the additive noise. The goal of graph filtering is to effectively recover the original signal $\mathbf{x}$ from the noisy observation $\mathbf{y}$.

\indent Instead of utilizing a single shared spectral response vector for the entire graph, we assign a customized filter to each vertex. To this end, we define a learnable dense coefficient matrix $\mathbf{H} \in \mathbb{R}^{N \times N}$, where the $i$-th row $\mathbf{H}_{i,:}$ encapsulates the exclusive spectral coefficients for the $i$-th node. Consequently, the filtering output for a specific node $i$ is obtained by applying its personalized spectral response to the transformed graph signal and extracting the $i$-th observation. Mathematically, this process can be expressed as
\begin{align}  \label{filter process}
	\tilde{x}_i &= \boldsymbol{\delta}_i^{\top} \mathcal{F}^{-1} \mathrm{diag}(\mathbf{H}_{i,:}) \mathcal{F} \mathbf{x} \nonumber\\
	&= \sum_{k=1}^{N} \mathcal{F}^{-1}_{i,k} H_{i,k} (\mathcal{F} \mathbf{x})_k,
\end{align}
where $\boldsymbol{\delta}_i \in \mathbb{R}^N$ denotes the $i$-th standard basis column vector.

\indent By generalizing this node-wise operation to the entire graph, the overall filtered signal $\mathbf{y}$ can be elegantly formulated in a compact matrix form using the Hadamard product $\odot$ as
\begin{align} 
	\tilde{\mathbf{x}} = \left( \mathcal{F}^{-1} \odot \mathbf{H} \right) \mathcal{F} \mathbf{x}.
\end{align}

\indent \emph{Remark 1:} When the $\mathcal{F}$ reduces to the GFT matrix, the proposed mechanism strictly reduces to the node-oriented spectral filtering method presented in \cite{ZhengNode24}. When $\mathbf{H}_{1,:} = \mathbf{H}_{2,:}= \cdots = \mathbf{H}_{N,:}$, the proposed mechanism naturally degrades to the traditional global fractional-domain graph filtering \cite{Ozturk21}.

\indent To optimize the filter coefficient matrix $\mathbf{H}$, we employ the mean squared error (MSE) as the objective function 
\begin{align}
	\mathcal{L}_{\mathrm{MSE}} = \frac{1}{N} \|\tilde{\mathbf{x}} - \mathbf{x}\|_2^2,
\end{align}
where $\|\cdot\|_2$ denotes the $\ell_2$-norm.

\indent Furthermore, the signal-to-noise ratio (SNR) provides a standard measure of the reconstruction fidelity and is formulated as
\begin{align}
	\mathrm{SNR} = 20 \log_{10} \frac{\|\mathbf{x}\|_2}{\|\tilde{\mathbf{x}} - \mathbf{x}\|_2}.
\end{align}

\subsection{Neural network architecture implementation}
\indent The proposed NOFF mechanism is implemented as an end-to-end trainable network composed of stacked node-oriented fractional filter layers.

\indent At the $l$-th layer, let $\mathbf{H}^{(l)} \in \mathbb{R}^{N \times N}$ denote the learnable coefficient matrix, where the $i$-th row $\mathbf{H}_{i,:}^{(l)}$ specifies the vertex-dependent spectral response of node $i$. Meanwhile, the forward and inverse generalized fractional-domain transforms are parameterized by layer-specific learnable variables, denoted by $\mathcal{F}^{(l)}$ and $(\mathcal{F}^{(l)})^{-1}$, respectively. Their exact forms depend on the selected generalized fractional-domain system in Section \ref{sec2}. Let $\mathbf{x}^{(0)}=\mathbf{y}$ denote the noisy input. Then the output of the $l$-th filter layer is given by
\begin{align}
\mathbf{z}^{(l+1)} = \left( (\mathcal{F}^{(l)})^{-1} \odot \mathbf{H}^{(l)} \right)\mathcal{F}^{(l)} \mathbf{x}^{(l)}.
\end{align}

\indent To improve representation capability and stabilize training, residual connections are introduced between consecutive layers. In this work, a three-layer architecture is adopted. For the first two layers, the hidden representations are updated by
\begin{align}
\mathbf{x}^{(l+1)} = \mathrm{ReLU}\Big(\Re(\mathbf{x}^{(l)}+\mathbf{z}^{(l+1)})\Big), \quad l\in\{0,1\},
\end{align}
where $\Re(\cdot)$ denotes the real-part operator, and $\mathrm{ReLU}(\cdot)$ denotes the rectified linear unit activation. In the last layer, the activation function is omitted to avoid truncating negative signal values, and the final reconstructed signal is obtained as
\begin{align}
\tilde{\mathbf{x}} = \mathbf{x}^{(2)} + \mathbf{z}^{(3)}.
\end{align}

\indent It is crucial to emphasize that the proposed NOFF is designed as a highly flexible and unified framework rather than being restricted to a single specific transform. We first clarify the terminology used throughout this paper to distinguish between different filtering paradigms. For any specific fractional transform, we evaluate both its global and node-oriented implementations. In the global fractional filtering (GFF) variants, abbreviated as GFF-[Transform], a single identical spectral response is shared across all vertices. In contrast, our proposed node-oriented fractional filtering variants, denoted as NOFF-[Transform], assign customized spectral responses to individual nodes. As systematically compared in Table \ref{tab:transform_comparison}, various fractional domains can be seamlessly instantiated within the proposed NOFF framework to address distinct physical challenges.

\begin{table*}[htbp]
	\centering
	\caption{Summary of implementations in the NOFF framework}
	\label{tab:transform_comparison}
	\begin{tabular}{lcccp{4.5cm}}
		\toprule
		\textbf{Method} & \textbf{Layer formulation $\mathbf{z}^{(l+1)}$} & \textbf{Filtering domain} & \textbf{Learnable parameters} & \textbf{Spectral Advantage \& Role} \\
		\midrule
		NOFF-GFT 
		& $\left( \mathbf{F}^{-1} \odot \mathbf{H}^{(l)} \right) \mathbf{F} \mathbf{x}^{(l)}$ 
		& GFT 
		& $\mathbf{H}$ 
		& Rigid spectral basis. Serves as a baseline, but lacks flexibility to decouple complex aliased noise. \\
		NOFF-GFRFT 
		& $\left( \left( (\mathbf{F}^{\alpha})^{-1}\right)^{(l)}  \odot \mathbf{H}^{(l)} \right) \left( \mathbf{F}^{\alpha}\right)^{(l)}  \mathbf{x}^{(l)}$ 
		& GFRFT 
		& $\mathbf{H}, \alpha$ 
		& Global fractional power. Effective for separating globally uniform noise via non-linear spectral scaling. \\
		NOFF-MPGFRFT-I 
		& $\left( \left( (\mathbf{F}_{\rm I}^{\boldsymbol{a}})^{-1}\right)^{(l)} \odot \mathbf{H}^{(l)} \right) \left( \mathbf{F}_{\rm I}^{\boldsymbol{a}}\right)^{(l)} \mathbf{x}^{(l)}$ 
		& MPGFRFT-I 
		& $\mathbf{H}, \boldsymbol{a}$ 
		& Independent fractional power per spectral component. Essential for decoupling multi-scale heterogeneous patterns. \\
		NOFF-MPGFRFT-II 
		& $\left( \left( (\mathbf{F}_{\rm II}^{\boldsymbol{a}})^{-1}\right)^{(l)} \odot \mathbf{H}^{(l)} \right) \left( \mathbf{F}_{\rm II}^{\boldsymbol{a}}\right)^{(l)} \mathbf{x}^{(l)}$ 
		& MPGFRFT-II 
		& $\mathbf{H}, \boldsymbol{a}$ 
		& Fine-grained modulation via matrix polynomials. Maximizes adaptive capacity to resolve highly localized anomalies. \\
		NOFF-Lap-GLCT 
		& $\left( \left( (\mathbf{F}_{\mathrm{I}}^{\mathbf{M}})^{-1}\right)^{(l)} \odot \mathbf{H}^{(l)} \right) \left( \mathbf{F}_{\mathrm{I}}^{\mathbf{M}}\right)^{(l)} \mathbf{x}^{(l)}$ 
		& Lap-GLCT
		& $\mathbf{H}, \mathbf{M}$ 
		& Laplacian-based affine parameterization. Specifically targets and compensates for topological spectral shifts. \\
		NOFF-wAdj-GLCT  
		& $\left( \left( (\mathbf{F}_{\mathrm{II}}^{\mathbf{M}})^{-1}\right)^{(l)} \odot \mathbf{H}^{(l)} \right) \left( \mathbf{F}_{\mathrm{II}}^{\mathbf{M}}\right)^{(l)} \mathbf{x}^{(l)}$
		& wAdj-GLCT 
		& $\mathbf{H}, \mathbf{M}$ 
		& Adjacency-based affine transformation. Uniquely suited to capture asymmetric signal propagation and complex structural dynamics. \\
		\bottomrule
	\end{tabular}
\end{table*}

\indent The overall training and inference procedure of the proposed network is summarized in Algorithm~\ref{alg:node_frac_training}. This localized architecture empowers the model to capture complex topological variations. 

\begin{algorithm}[!t]
\caption{Training and inference of the proposed NOFF network}
\label{alg:node_frac_training}
\begin{algorithmic}[1]
\REQUIRE Training set $\mathcal{D}_{\mathrm{tr}}$, validation set $\mathcal{D}_{\mathrm{val}}$, test set $\mathcal{D}_{\mathrm{te}}$, GSO $\mathbf{S}$, maximum epoch $E$, learning rate $\eta$
\ENSURE Learned filter matrices $\{\mathbf{H}^{(l)}\}_{l=0}^{2}$, learned layer-specific transform parameters, and the final filtered signal $\tilde{\mathbf{x}}$
\STATE \textbf{Initialization:} Compute the transform basis from $\mathbf{S}$ and initialize filter matrices $\mathbf{H}^{(l)}_{{l=0}^{2}}$ and layer-specific transform parameters
\FOR{each epoch $e=1,2,\ldots,E$}
    \FOR{each training sample pair $(\mathbf{y},\mathbf{x})\in\mathcal{D}_{\mathrm{tr}}$}
        \STATE Set $\mathbf{x}^{(0)}\leftarrow \mathbf{y}$
        \FOR{$l=0,1,2$}
            \STATE Compute $\mathbf{z}^{(l+1)}=\big((\mathcal{F}^{(l)})^{-1}\odot\mathbf{H}^{(l)}\big)\mathcal{F}^{(l)}\mathbf{x}^{(l)}$
            \IF{$l<2$}
                \STATE Update $\mathbf{x}^{(l+1)}\leftarrow \mathrm{ReLU}(\Re(\mathbf{x}^{(l)}+\mathbf{z}^{(l+1)}))$
            \ENDIF
        \ENDFOR
        \STATE Obtain $\tilde{\mathbf{x}}\leftarrow \mathbf{x}^{(2)}+\mathbf{z}^{(3)}$
        \STATE \textbf{Loss Computation \& Optimization:} Compute $\mathcal{L}_{\mathrm{MSE}}=\|\tilde{\mathbf{x}}-\mathbf{x}\|_2^2$ and update $\{\mathbf{H}^{(l)},\mathcal{F}^{(l)}\}_{l=0}^{2}$ using Adam
    \ENDFOR
    \STATE After each epoch, evaluate the validation loss on $\mathcal{D}_{\mathrm{val}}$
	\STATE Update the learning rate scheduler and early-stopping state according to the validation loss
\ENDFOR
\STATE Load the best saved network parameters and reconstruct the signals in $\mathcal{D}_{\mathrm{te}}$
\STATE Compute the test SNR and report
\end{algorithmic}
\end{algorithm}

\indent However, it is important to acknowledge the inherent limitations of this ideal formulation. While the unconstrained NOFF theoretically provides exceptional spatial-spectral adaptability, its direct application to large-scale graphs encounters a severe curse of dimensionality. For a graph with $N$ vertices, the unconstrained filter matrix $\mathbf{H}$ requires $N^2$ independent spectral parameters per layer. When $N$ becomes massive, this excessively vast parameter space not only incurs prohibitive memory and computational costs but also introduces a severe risk of overfitting, leading the model to memorize random graph noise. Therefore, it is imperative to structurally regularize the parameter space to handle large-scale networks effectively. We will address this critical computational challenge in the following section.

\section{Scalable Node-Oriented Fractional Filtering via Low-Rank Approximation}  \label{sec4}
\indent To overcome the curse of dimensionality and the risk of overfitting inherent in the unconstrained NOFF model, this section proposes a scalable LRNOFF architecture. By imposing a structural bottleneck, we drastically reduce the parameter space while preserving the adaptability.

\subsection{Low-rank formulation and efficient computation}
\indent To mitigate this scalability bottleneck, we introduce a low-rank approximation algorithm for the dense coefficient matrix $\mathbf{H}$. Specifically, we constrain $\mathbf{H}$ by decomposing it into the product of two low-rank sub-matrices:
\begin{align} \label{eq:low_rank_H}
	\mathbf{H} = \mathbf{W}\mathbf{B},
\end{align}
where the spatial weight matrix $\mathbf{W} \in \mathbb{R}^{N \times d}$ and the base spectral filter matrix $\mathbf{B} \in \mathbb{R}^{d \times N}$, with the rank satisfying $d \ll N$. Consequently, the filtering coefficient for the $i$-th node and $j$-th spectral component is given by $H_{i,j} = \sum_{k=1}^{d} W_{i,k} B_{k,j}$.

\indent Without the low-rank structure, computing the naive node-oriented filtering directly from its definition requires executing the spectral response mapping for each vertex independently, which yields a complexity of $\mathcal{O}(N^3)$. Such an extensive computational burden is impractical for large-scale graph analysis. However, by substituting the low-rank constraint into the filtering equation, the overall computation can be elegantly decoupled and accelerated. We formalize this efficient computational process in the following theorem.

\indent \emph{Theorem 1:} Given the forward and inverse fractional-domain transform matrices $\mathcal{F}$ and $\mathcal{F}^{-1}$, the input signal $\mathbf{x}$, and the low-rank filter decomposition $\mathbf{H} = \mathbf{W}\mathbf{B}$ with rank $d$, the filtered output $\tilde{\mathbf{x}} = \left( \mathcal{F}^{-1} \odot \mathbf{H} \right) \mathcal{F} \mathbf{x}$ can be equivalently computed as
\begin{align} \label{eq:theorem_fast_compute}
	\tilde{\mathbf{x}} = \sum_{k=1}^{d} \left[ \mathbf{w}_k \odot \left( \mathcal{F}^{-1} (\mathbf{b}_k \odot \mathcal{F}\mathbf{x}) \right) \right],
\end{align}
where $\mathbf{w}_k = \mathbf{W}_{:,k} \in \mathbb{C}^N$ represents the $k$-th column of $\mathbf{W}$, and $\mathbf{b}_k = (\mathbf{B}_{k,:})^{\top} \in \mathbb{C}^N$ represents the transpose of the $k$-th row of $\mathbf{B}$.

\indent \emph{Proof:} See Appendix \ref{Appendix A}.

\indent \emph{Remark 2 (Complexity analysis):} Before evaluating the computational efficiency, it is crucial to clarify that the construction of bases $\mathcal{F}$ and $\mathcal{F}^{-1}$) is assumed to be pre-computed as a standard pre-processing step in GSP. We focus strictly on the matrix calculation complexity of the filtering operation itself. 
\begin{itemize}
	\item \textbf{GFF:} Since all nodes share a single spectral response vector, the operation involves one forward transform, one diagonal multiplication, and one inverse transform, resulting in a matrix calculation complexity of $\mathcal{O}(N^2)$.
	\item \textbf{Unconstrained NOFF:} Without the low-rank constraint, each node independently learns its own response. The unconstrained dense matrix operations yield a complexity of $\mathcal{O}(N^3)$.
	\item \textbf{Low-rank NOFF via direct computation (LRNOFF-Direct):} Even if the low-rank constraint $\mathbf{H} = \mathbf{W}\mathbf{B}$ is applied, computing the filter directly without the summation rearrangement of Theorem 1 requires evaluating the coupled dense matrix operations. This still incurs a complexity of $\mathcal{O}(N^3)$, which fails to resolve the computational bottleneck.
	\item \textbf{Low-rank NOFF via Theorem 1 (LRNOFF-Fast):} As demonstrated in Theorem 1, the computation is decoupled into $d$ parallel fractional transform operations. This allows the matrix calculation complexity to drop significantly to $\mathcal{O}(dN^2)$.
\end{itemize}

\indent \emph{Remark 3 (Physical interpretation):} From a physical perspective, the decomposition $\mathbf{H}=\mathbf{W}\mathbf{B}$ provides a meaningful interpretation of the filtering process. The matrix $\mathbf{B}$ acts as a \emph{global dictionary} that learns $d$ fundamental fractional-spectral filtering patterns across the entire graph. Meanwhile, $\mathbf{W}$ functions as an \emph{adaptive allocator} that assigns localized weights for each vertex, smoothly blending these base patterns to accommodate the distinct topological neighborhood of each node.

\indent To systematically summarize the architectural evolution and the aforementioned advantages, Table \ref{tab:Comprehensive comparison} provides a comprehensive comparison of different filtering paradigms. It is evident that the proposed LRNOFF-Fast perfectly balances localized spatial adaptability, implicit regularization, and strict computational efficiency, rendering it highly scalable for large-scale graphs.

\begin{table*}[htbp]
	\centering
	\caption{Comprehensive comparison of theoretical attributes among various filtering paradigms}
	\label{tab:Comprehensive comparison}
	\resizebox{\textwidth}{!}{%
		\begin{tabular}{lccccc}
			\toprule
			\textbf{Method} & \textbf{Spatial adaptability} & \textbf{Implicit regularization} & \textbf{Filter parameters} & \textbf{Computational complexity} & \textbf{Large-scale feasibility} \\
			\midrule
			GFF 
			& Global 
			& $\times$ 
			& $\mathcal{O}(N)$ 
			& $\mathcal{O}(N^2)$
			& \checkmark \\
			
			Unconstrained NOFF 
			& Node-oriented 
			& $\times$ 
			& $\mathcal{O}(N^2)$ 
			& $\mathcal{O}(N^3)$
			& $\times$ \\
			
			LRNOFF-Direct 
			& Node-oriented 
			& \checkmark (Low-rank) 
			& $\mathcal{O}(dN)$ 
			& $\mathcal{O}(N^3)$
			& \checkmark \\
			
			LRNOFF-Fast 
			& Node-oriented 
			& \checkmark (Low-rank) 
			& $\mathcal{O}(dN)$ 
			& $\mathcal{O}(dN^2)$
			& \checkmark \\
			\bottomrule
		\end{tabular}%
	}
\end{table*}

\indent The closed-form optimal solution for GFF has been rigorously established by Ozturk et al. \cite{Ozturk21}. By constraining the fractional filter to a global diagonal matrix parameterized by a single response vector $\mathbf{h}_{opt} \in \mathbb{C}^{N \times 1}$, the optimal global filter coefficients are obtained by solving the linear system $\mathbf{T}\mathbf{h}_{opt} = \mathbf{q}$. However, achieving optimal node-oriented filtering requires removing this rigid global sharing constraint and assigning an independent filtering response to each vertex. The ideal unconstrained local filter vector for the $i$-th vertex, denoted as $\mathbf{h}_{opt}^{(i)} \in \mathbb{C}^{N \times 1}$, can be derived by individually minimizing its localized reconstruction MSE.

\indent By vertically concatenating these optimal unconstrained responses of all $N$ vertices, we can explicitly construct the complete, ideal node-oriented fractional filter matrix 
\begin{align} \label{eq:H_opt_ideal}
	\mathbf{H}_{opt} = \left[ \mathbf{h}_{opt}^{(1)}, \mathbf{h}_{opt}^{(2)}, \cdots, \mathbf{h}_{opt}^{(N)} \right]^{\top} \in \mathbb{C}^{N \times N}.
\end{align}

\subsection{Theoretical analysis and optimization of the LRNOFF-Fast}
\indent The proposed LRNOFF-Fast is implemented as an end-to-end trainable framework, sharing a similar architectural philosophy with the NOFF introduced in Section \ref{sec3}. Let $\mathcal{L}$ denote the MSE loss function and $\eta$ denote the learning rate. During the backpropagation process, the network parameters are updated iteratively via gradient descent. The complete gradient of the loss function with respect to all learnable parameters is defined as
\begin{align}
	\nabla \mathcal{L} = \left\{ \frac{\partial \mathcal{L}}{\partial \mathbf{W}}, \frac{\partial \mathcal{L}}{\partial \mathbf{B}}, \frac{\partial \mathcal{L}}{\partial \boldsymbol{\phi}} \right\}.
\end{align}

\indent At the $t$-th iteration, the spatial weight matrix $\mathbf{W}$, the base spectral filter matrix $\mathbf{B}$, and the fractional-domain parameter $\boldsymbol{\phi}$ (e.g., the fractional order $\alpha$ in GFRFT, the fractional order vector $\boldsymbol{a}$ in MPGFRFT, or the matrix $\mathbf{M}$ in GLCT) are updated by
\begin{align}
	\mathbf{W}^{(t+1)} = \mathbf{W}^{(t)} - \eta \frac{\partial \mathcal{L}}{\partial \mathbf{W}^{(t)}},\\
	\mathbf{B}^{(t+1)} = \mathbf{B}^{(t)} - \eta \frac{\partial \mathcal{L}}{\partial \mathbf{B}^{(t)}},\\
	\boldsymbol{\phi}^{(t+1)} = \boldsymbol{\phi}^{(t)} - \eta \frac{\partial \mathcal{L}}{\partial \boldsymbol{\phi}^{(t)}}.
\end{align}

\indent For the spatial and spectral filter components, the gradients drive the model to learn the optimal localized blending and global base patterns, respectively. Since the generalized fractional transforms are fully differentiable with respect to their defining variables, the optimization of the fractional parameter $\boldsymbol{\phi}$ is directly handled through standard automatic differentiation to dynamically search for the optimal transform domain. By the fundamental definition of partial differentiation, computing the gradient for one specific parameter matrix inherently treats the remaining variables as constants. To provide a rigorous theoretical guarantee for this gradient-based learning process, the following two theorems demonstrate that under such conditional fixation, the separate optimizations of $\mathbf{W}$ and $\mathbf{B}$ correspond directly to deriving exact, closed-form minimum MSE (MMSE) solutions.

\indent \emph{Theorem 2:} Assume the base spectral filter matrix $\mathbf{B}$ is fixed, and let $\mathbf{u}_k$ denote the $k$-th base filtered signal mapped to the vertex domain. For the $i$-th vertex, let $\mathbf{u}^{(i)} = [U_{i,1}, U_{i,2}, \cdots, U_{i,d}]^{\top}$, and $\mathbf{w}^{(i)} = [W_{i,1}, W_{i,2}, \cdots, W_{i,d}]^{\top}$. By minimizing the local MSE cost function $J(\mathbf{w}^{(i)}) = \mathbb{E}[|\tilde{x}_i - x_i|^2]$, the optimal spatial weight vector is derived as
\begin{align} \label{eq:theorem2_W}
	\mathbf{w}_{opt}^{(i)} = (\mathbf{R}_{U}^{(i)})^{-1} \mathbf{r}_{Ux}^{(i)}, \quad \forall i \in \{1, 2, \cdots, N\},
\end{align}
where $\mathbf{R}_{U}^{(i)} = \mathbb{E}[(\mathbf{u}^{(i)})^* (\mathbf{u}^{(i)})^{\top}] \in \mathbb{C}^{d \times d}$ is the local auto-correlation matrix, and $\mathbf{r}_{Ux}^{(i)} = \mathbb{E}[(\mathbf{u}^{(i)})^* x_i] \in \mathbb{C}^{d \times 1}$ is the cross-correlation vector. The superscript $*$ denotes the complex conjugate.

\indent \emph{Proof:} See Appendix \ref{Appendix B}.

\indent \emph{Theorem 3:} Assume the spatial weight matrix $\mathbf{W}$ is fixed, and let $\mathbf{\Psi} = \mathcal{F}^{-1} \mathrm{diag}(\mathcal{F}\mathbf{y}) \in \mathbb{C}^{N \times N}$ be the combined transform matrix for the input signal $\mathbf{y}$. By vectorizing the base filter matrix as a global column vector $\mathbf{b} = [(\mathbf{b}^{(1)})^{\top}, (\mathbf{b}^{(2)})^{\top}, \cdots, (\mathbf{b}^{(d)})^{\top}]^{\top} \in \mathbb{C}^{dN \times 1}$ where $\mathbf{b}^{(k)} = (\mathbf{B}_{k,:})^{\top}$, and defining a concatenated spatial-spectral feature row vector for node $i$ as $\mathbf{v}^{(i)} = [W_{i,1}\mathbf{\Psi}_{i,:}, \cdots, W_{i,d}\mathbf{\Psi}_{i,:}] \in \mathbb{C}^{1 \times dN}$, the optimal global base filter vector $\mathbf{b}$ that minimizes the global MSE $J(\mathbf{b}) = \sum_{i=1}^N \mathbb{E}[|\mathbf{v}^{(i)} \mathbf{b} - x_i|^2]$ is given by
\begin{align} \label{eq:theorem3_B}
	\mathbf{b}_{opt} = \mathbf{R}_{V}^{-1} \mathbf{r}_{Vx},
\end{align}
where $\mathbf{R}_{V} = \sum_{i=1}^{N} \mathbb{E}[(\mathbf{v}^{(i)})^{\mathrm{H}} \mathbf{v}^{(i)}] \in \mathbb{C}^{dN \times dN}$ is the global auto-correlation matrix, and $\mathbf{r}_{Vx} = \sum_{i=1}^{N} \mathbb{E}[(\mathbf{v}^{(i)})^{\mathrm{H}} x_i] \in \mathbb{C}^{dN \times 1}$ is the global cross-correlation vector.

\indent \emph{Proof:} See Appendix \ref{Appendix C}.

\indent \emph{Remark 3 (Structural disparity in objective functions):} It is crucial to note the fundamental discrepancy between the objective functions employed in Theorems 2 and 3. Since the spatial weight matrix $\mathbf{W}$ assigns an independent vector to each specific node, its optimization can be mathematically decoupled into $N$ independent local MMSE problems. In contrast, the base spectral filter matrix $\mathbf{B}$ acts as a globally shared dictionary across the entire graph. Consequently, the parameters in $\mathbf{B}$ cannot be decoupled node-by-node; their optimization must be formulated as a joint global MMSE problem that minimizes the aggregated reconstruction error over all vertices simultaneously.

\indent \emph{Remark 4 (Theoretical guarantee for gradient-based learning):} Theorems 2 and 3 provide a rigorous statistical signal processing foundation for the iterative gradient descent updates of the network. In practice, the model dynamically updates the parameters via partial derivatives (i.e., descending along $\frac{\partial \mathcal{L}}{\partial \mathbf{W}}$ and $\frac{\partial \mathcal{L}}{\partial \mathbf{B}}$) rather than directly calculating the computationally prohibitive matrix inversions required by the exact Wiener solutions. However, these theorems mathematically guarantee that the conditional loss landscape for each parameter block is strictly convex. Consequently, the partial gradients computed during backpropagation are theoretically well-posed, stably guiding the optimization trajectory toward the exact closed-form MMSE optimums without the risk of gradient divergence.

\indent To establish the structural validity of our proposed low-rank architecture, the following theorem demonstrates how this massive, parameter-heavy ideal filter matrix $\mathbf{H}_{opt}$ can be optimally approximated.

\indent \emph{Theorem 4:} Assume the ideal unconstrained filter matrix $\mathbf{H}_{opt}$ defined in Eq. (\ref{eq:H_opt_ideal}) admits the singular value decomposition (SVD) $\mathbf{H}_{opt} = \mathbf{U} \mathbf{\Sigma} \mathbf{V}^{\mathrm{H}}$. Under the low-rank constraint $\mathrm{rank}(\mathbf{H}) \le d$, the global optimal matrix $\mathbf{H}_{d}^{\star}$ that minimizes the Frobenius norm approximation error $\|\mathbf{H}_{opt} - \mathbf{H}\|_F^2$ is exactly given by the truncated SVD
\begin{align} \label{eq:theorem4_svd}
	\mathbf{H}_{d}^{\star} = \mathbf{U}_d \mathbf{\Sigma}_d \mathbf{V}_d^{\mathrm{H}},
\end{align}
where $\mathbf{U}_d \in \mathbb{C}^{N \times d}$ and $\mathbf{V}_d \in \mathbb{C}^{N \times d}$ contain the first $d$ left and right singular vectors, respectively, and $\mathbf{\Sigma}_d \in \mathbb{R}^{d \times d}$ is the diagonal matrix of the $d$ largest singular values. Furthermore, this optimal projection can be symmetrically factorized into a localized spatial weight matrix $\mathbf{W}_{opt} \in \mathbb{C}^{N \times d}$ and a global base spectral filter matrix $\mathbf{B}_{opt} \in \mathbb{C}^{d \times M}$, given by 
\begin{align}
	\mathbf{W}_{opt} = \mathbf{U}_d \mathbf{\Sigma}_d^{1/2}, \\
	\mathbf{B}_{opt} = \mathbf{\Sigma}_d^{1/2} \mathbf{V}_d^{\mathrm{H}}.
\end{align}

\indent \emph{Proof:} See Appendix \ref{Appendix D}.

\indent \emph{Remark 5 (Bridging analytical SVD projection with implicit regularization):} Theorem 4 mathematically justifies our decoupled $\mathbf{W}\mathbf{B}$ architecture. It proves that while the ideal unconstrained filter $\mathbf{H}_{opt}$ requires a massive parameter space, its most essential and dominant spectral filtering properties can be optimally captured and bounded within a low-rank subspace governed by the principal singular values. In practical large-scale applications, rather than directly calculating the highly complex $\mathbf{H}_{opt}$, which would easily lead to severe overfitting on random noise, our proposed LRNOFF-Fast adopts this theoretically guaranteed $\mathbf{W}\mathbf{B}$ structure as a hypothesis space. By dynamically optimizing $\mathbf{W}$ and $\mathbf{B}$ via end-to-end MSE minimization, the low-rank constraint ($d \ll N$) acts as a powerful implicit regularizer. It strictly forces the network to learn the noise-free principal filter bases (analogous to $\mathbf{B}_{opt}$) and their spatial combinations (analogous to $\mathbf{W}_{opt}$), perfectly aligning the mathematical optimum with robust, data-driven empirical performance.

\section{Experiments} \label{sec5} 
\subsection{Experiments on small-scale graphs} \label{sec5.3}
\indent \textbf{Datasets and Graph Construction:} To evaluate denoising performance on graph signals, we consider two representative real-world datasets, namely the Quality dataset ($N = 7$) \cite{Kun23Frequency} and the Exchange-rate dataset ($N = 8$) \cite{lai2018modeling}. For each dataset, we use symmetric adjacency matrix $\mathbf{A}_{\mathrm{sym}} = (\mathbf{A} + \mathbf{A}^{\top})/2$ and the normalized symmetric graph Laplacian $\mathbf{L}_{\mathrm{sym}} = \mathbf{I} - \mathbf{D}^{-1/2} \mathbf{A}_{\mathrm{sym}} \mathbf{D}^{-1/2}$. The eigendecomposition of $\mathbf{L}_{\mathrm{sym}}$ then provides the orthogonal graph spectral basis used to construct the generalized transform basis $\mathcal{F}$. In this way, all compared methods are evaluated on the same graph topology and differ only in the adopted transform/filter design.

\indent \textbf{Setup:} To construct the data samples, we extract the first $1500$ temporal observations from each dataset. The truncated temporal sequence is then reshaped into graph signal snapshots and split chronologically into training, validation, and testing subsets with a ratio of 60\%, 20\%, and 20\%, respectively. All models are trained using Adam with an initial learning rate of 0.001 and weight decay of 0.001. The optimization objective is the MSE loss. We train for at most 500 epochs, reduce the learning rate by a factor of 0.5 when the validation loss does not improve for 10 epochs, and employ early stopping with a patience of 30 epochs. The model achieving the lowest validation loss is restored for testing, and the final reconstruction quality is reported in terms of SNR on the unseen test set.

\indent To highlight the benefit of our proposed NOFF mechanism, we organize the baselines into matched pairs. Specifically, global fractional filtering (GFF) methods share a single common filter across the entire graph, whereas NOFF methods assign an independent spectral filtering response to each node. This paired design allows us to isolate the performance gain brought by the node-oriented spatial adaptation from the gain brought by the fractional transform family itself. For fair comparison, all methods are initialized under the same principle. The filter matrix is initialized as an all-ones matrix. For GFRFT-based methods, the fractional order is initialized as $\alpha=1$. For MPGFRFT-based methods, the order vector is initialized as $\boldsymbol{a}=(1,1,\cdots,1)$.  Table~\ref{tab:global_vs_node_small} first compares global and node-oriented filtering under several fractional-domain transforms, including GFRFT, MPGFRFT-I, MPGFRFT-II, Lap-GLCT and wAdj-GLCT. For simplicity of presentation in the tables, we abbreviate the global and node-oriented variants of a specific transform as GFF-[Transform] and NOFF-[Transform], respectively. It is evident that the node-oriented methods consistently outperform their global counterparts across all evaluated fractional domains. This performance gap strongly validates the necessity of our proposed node-level spatial adaptation. It fully unleashes the potential of fractional transforms, allowing for localized and proactive spectral modulation that a globally shared filter is fundamentally incapable of achieving.

\begin{table*}[htbp]
	\centering
	\caption{Denoising performance comparison of GFF-based and NOFF-based methods on small-scale datasets.}
	\label{tab:global_vs_node_small}
		\begin{tabular}{lcccccccc}
			\toprule
			Dataset & \multicolumn{4}{c}{Quality} & \multicolumn{4}{c}{Exchange-rate} \\
			\cmidrule(lr){2-5} \cmidrule(lr){6-9}
			$\sigma$ & $50$ & $80$ & $150$ & $200$ & $0.5$ & $0.8$ & $1$ & $1.2$ \\
			SNR & $3.949$ & $-0.133$ & $-5.593$ & $-8.092$ & $4.209$ & $0.126$ & $-1.812$ & $-3.395$ \\ \midrule
			GFF-GFRFT & $10.230$ & $8.094$ & $5.471$ & $4.841$ & $11.973$ & $8.581$ & $7.599$ & $7.428$ \\
			NOFF-GFRFT & $\mathbf{10.616}$ & $\mathbf{8.388}$ & $\mathbf{6.045}$ & $\mathbf{5.410}$ & $\mathbf{13.202}$ & $\mathbf{11.478}$ & $\mathbf{10.940}$ & $\mathbf{10.657}$ \\ \midrule
			GFF-MPGFRFT-I & $10.536$ & $8.304$ & $5.878$ & $5.169$ & $13.326$ & $11.137$ & $10.325$ & $8.975$ \\
			NOFF-MPGFRFT-I & $\mathbf{10.604}$ & $\mathbf{8.346}$ & $\mathbf{5.930}$ & $\mathbf{5.388}$ & $\mathbf{13.804}$ & $\mathbf{11.864}$ & $\mathbf{11.420}$ & $\mathbf{10.991}$ \\
			GFF-MPGFRFT-II & $10.421$ & $7.879$ & $5.492$ & $4.605$ & $10.168$ & $11.351$ & $10.869$ & $8.936$ \\
			NOFF-MPGFRFT-II & $\mathbf{10.422}$ & $\mathbf{8.376}$ & $\mathbf{5.812}$ & $\mathbf{4.868}$ & $\mathbf{13.929}$ & $\mathbf{11.842}$ & $\mathbf{11.317}$ & $\mathbf{10.460}$ \\ \midrule
			GFF-Lap-GLCT & $10.272$ & $8.184$ & $5.870$ & $4.688$ & $11.855$ & $8.672$ & $7.585$ & $7.786$ \\
			NOFF-Lap-GLCT & $\mathbf{10.471}$ & $\mathbf{8.272}$ & $\mathbf{5.933}$ & $\mathbf{5.428}$ & $\mathbf{13.187}$ & $\mathbf{11.455}$ & $\mathbf{10.878}$ & $\mathbf{10.683}$ \\
			GFF-wAdj-GLCT & $10.488$ & $8.117$ & $5.394$ & $4.613$ & $13.321$ & $9.832$ & $9.733$ & $9.275$ \\
			NOFF-wAdj-GLCT & $\mathbf{10.513}$ & $\mathbf{8.336}$ & $\mathbf{5.994}$ & $\mathbf{5.224}$ & $\mathbf{13.442}$ & $\mathbf{11.512}$ & $\mathbf{11.139}$ & $\mathbf{10.838}$ \\
			\bottomrule
		\end{tabular}%
\end{table*}

\begin{figure*}[htbp]
	\centering
	\includegraphics[width=\textwidth]{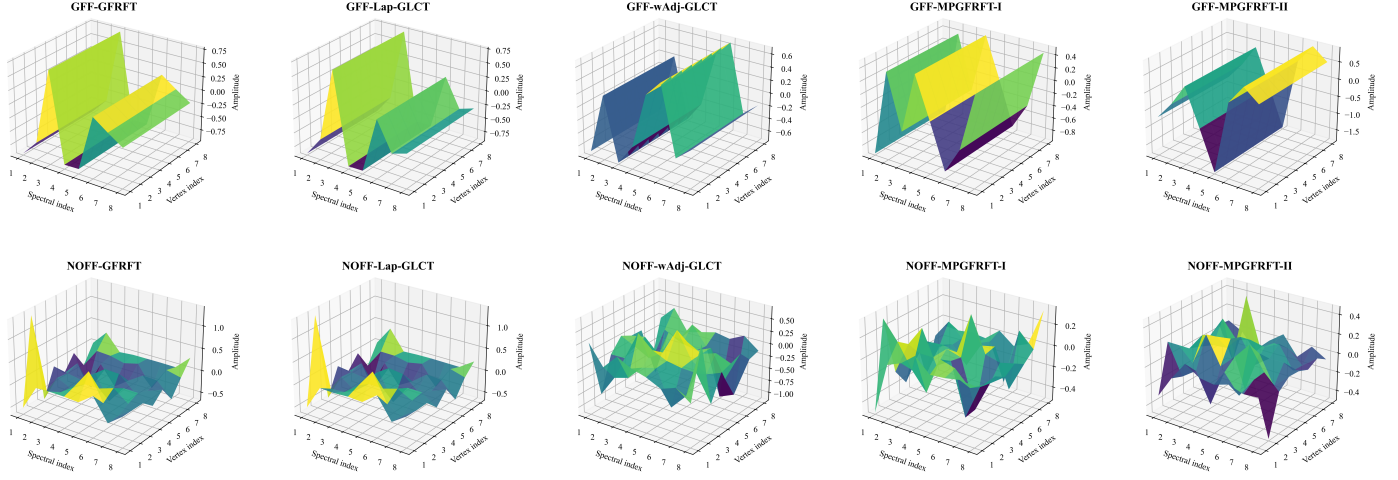} 
	\caption{Visualization of the learned 3D filter surfaces on the Exchange-rate dataset ($\sigma=0.5$, Layer 1).}
	\label{fig:learned_filters}
\end{figure*}

\begin{figure}[htbp]
	\centering
	\subfloat[]{
		\includegraphics[width=0.45\linewidth]{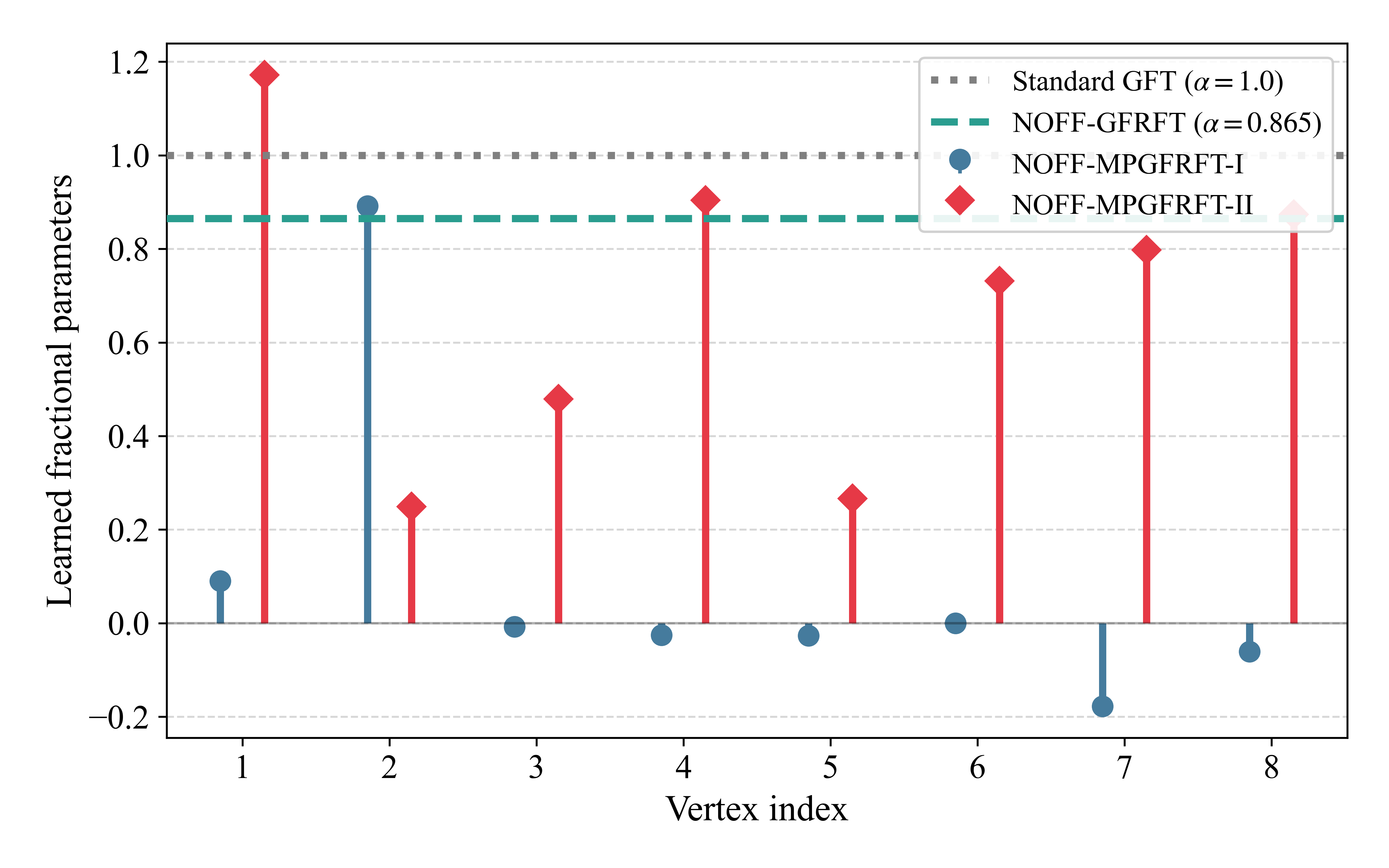}%
		\label{fig:param_a}%
	}%
	\hfill 
	\subfloat[]{
		\includegraphics[width=0.45\linewidth]{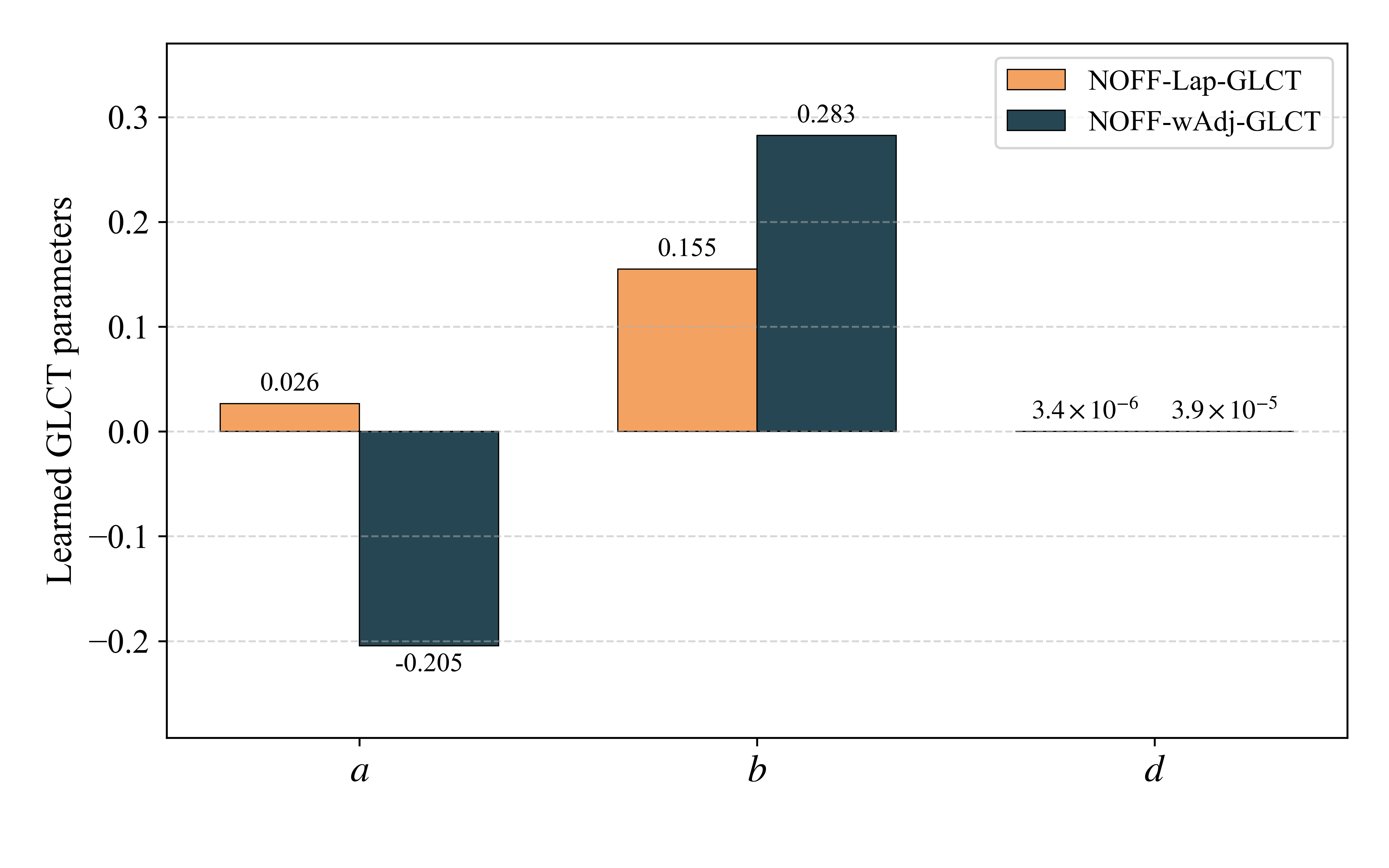}%
		\label{fig:param_b}%
	}
	
	\caption{Distribution of the optimized fractional parameters on the Exchange-rate dataset ($\sigma=0.5$, Layer 1). (a) Learned parameters for GFRFT and MPGFRFT variants. (b) Learned phase parameters ($a, b, d$) for GLCT variants.}
	\label{fig:learned_parameters}
\end{figure}

\indent To provide deeper physical insights into the superior performance of the NOFF framework, we visualize the learned spectral filters and fractional parameters. All visualizations are extracted from the first layer of the models evaluated on the Exchange-rate dataset under a noise level of $\sigma=0.5$. Fig. \ref{fig:learned_filters} illustrates the 3D surfaces of the learned filters across five fractional domains.  In the top row, the GFF-methods exhibit invariant spectral responses along the vertex index axis, confirming that a rigid, one-size-fits-all profile is enforced across the entire graph. In stark contrast, the bottom row reveals that the proposed NOFF framework generates highly rugged and localized filter surfaces. The spectral response dynamically adapts to the specific local topological environment of each vertex, providing compelling visual evidence for the necessity of node-oriented spatial adaptation. Fig. \ref{fig:learned_parameters} plots the distribution of the optimized fractional parameters. Fig. \ref{fig:learned_parameters}\subref{fig:param_a} shows the learned orders for GFRFT and MPGFRFT variants. It is evident that all learned parameters systematically deviate from the standard GFT domain ($\alpha=1.0$). Notably, the MPGFRFT variants assign highly distinctive fractional orders to individual vertices, fully exploiting their multi-parameter degrees of freedom. Similarly, Fig. \ref{fig:learned_parameters}\subref{fig:param_b} demonstrates that the learned phase parameters ($a, b, d$) for the GLCT variants also converge to non-trivial optimal values. These parameter deviations provide explicit empirical validation that our framework successfully locates an optimal intermediate transform space, thereby decoupling complex noise components that remain severely entangled in the standard spectral domain.

\indent To further demonstrate the superiority of the proposed framework, we extract the best-performing node-oriented configurations from the fractional domains (i.e., the optimal variants from GFRFT, MPGFRFT, and GLCT) and compare them against the node-oriented GFT (NO-GFT) filter as well as several state-of-the-art GNN and graph transformer methods. As reported in Table \ref{tab:fractional_vs_others_small}, the proposed NOFF methods demonstrate a clear and consistent advantage over the NO-GFT approach across all evaluated scenarios. This observation perfectly corroborates our theoretical motivation: although the node-oriented GFT introduces spatial adaptability, its reliance on the rigidly fixed standard spectral domain restricts its capacity to decouple the true signal from complex noise. In contrast, the fractional domains provide crucial extra degrees of freedom, enabling proactive spectral modulation to separate aliased frequency components. Meanwhile, they also achieve stronger denoising performance than the considered neural baselines, further confirming the effectiveness and competitiveness of the proposed fractional-domain framework.

\begin{table*}[htbp]
	\centering
	\caption{Denoising performance comparison of NOFF and state-of-the-art baselines on small-scale datasets.}
	\label{tab:fractional_vs_others_small}
		\begin{tabular}{lcccccccc}
			\toprule
			Dataset & \multicolumn{4}{c}{Quality} & \multicolumn{4}{c}{Exchange-rate} \\
			\cmidrule(lr){2-5} \cmidrule(lr){6-9}
			$\sigma$ & $50$ & $80$ & $150$ & $200$ & $0.5$ & $0.8$ & $1$ & $1.2$ \\
			SNR & $3.949$ & $-0.133$ & $-5.593$ & $-8.092$ & $4.209$ & $0.126$ & $-1.812$ & $-3.395$ \\
			\midrule
			APPNP & $7.078$ & $4.963$ & $3.521$ & $3.250$ & $7.781$ & $6.179$ & $5.728$ & $5.466$ \\
			GAT & $7.088$ & $4.996$ & $3.578$ & $3.268$ & $7.542$ & $6.070$ & $5.644$ & $5.409$ \\
			GCN & $7.006$ & $4.945$ & $3.527$ & $3.254$ & $7.626$ & $6.111$ & $5.678$ & $5.432$ \\
			H2GCN & $8.180$ & $5.984$ & $3.815$ & $3.159$ & $7.757$ & $5.888$ & $5.252$ & $4.857$ \\
			ChebyNet & $7.120$ & $5.013$ & $3.560$ & $3.265$ & $7.942$ & $6.315$ & $5.855$ & $5.582$ \\
			BernNet & $6.761$ & $4.446$ & $2.693$ & $2.259$ & $6.698$ & $4.701$ & $4.052$ & $3.637$ \\
			GPR-GNN & $8.111$ & $5.950$ & $3.770$ & $3.307$ & $9.363$ & $7.571$ & $7.015$ & $6.689$ \\
			ARMAConv & $7.133$ & $5.019$ & $3.591$ & $3.279$ & $7.815$ & $6.221$ & $5.764$ & $5.497$ \\
			SpecFormer & $7.516$ & $5.768$ & $4.416$ & $4.079$ & $10.436$ & $9.269$ & $8.924$ & $8.717$ \\
			UniMP & $7.114$ & $5.028$ & $3.550$ & $3.279$ & $7.882$ & $6.294$ & $5.838$ & $5.559$ \\
			PolyFormer & $7.619$ & $5.244$ & $3.711$ & $3.199$ & $7.389$ & $5.970$ & $5.567$ & $5.340$ \\
			\midrule
			NO-GFT & $10.425$ & $8.124$ & $5.833$ & $5.119$ & $12.879$ & $10.927$ & $10.270$ & $10.017$ \\
			NOFF-GFRFT & $\mathbf{10.616}$ & $\mathbf{8.388}$ & $\mathbf{6.045}$ & $5.410$ & $13.202$ & $11.478$ & $10.940$ & $10.657$ \\
			NOFF-MPGFRFT & $10.604$ & $8.376$ & $5.930$ & $5.388$ & $\mathbf{13.929}$ & $\mathbf{11.864}$ & $\mathbf{11.420}$ & $\mathbf{10.991}$ \\
			NOFF-GLCT & $10.513$ & $8.336$ & $5.994$ & $\mathbf{5.428}$ & $13.442$ & $11.512$ & $11.139$ & $10.838$ \\
			\bottomrule
		\end{tabular}%
\end{table*}

\subsection{Experiments on large-scale graphs}
\indent \textbf{Datasets and Graph Construction:} To evaluate the scalability and denoising performance of the proposed LRNOFF mechanism on large-scale networks, we conduct experiments on four widely-used real-world spatiotemporal datasets: METR ($N=207$), PEMS08 ($N=170$) \cite{yan2026jfrffnet,li2018diffusion,Guo22Learning}, Solar ($N=137$) \cite{nrel_solar_data} and Electricity ($N=370$) \cite{Lai18Modeling}. Specifically, to ensure temporally aligned dense observations for the Electricity dataset, we truncate the initial inactive periods and construct a symmetric $5$-nearest neighbor ($5$-NN) graph based on the Pearson correlation of average daily load profiles. For the Solar dataset, lacking spatial coordinates, its symmetric $5$-NN graph topology is similarly built using the positive Pearson correlations of the full time series. Finally, for all datasets, the graph shift operator and the corresponding generalized transform bases are constructed following the same normalized symmetric Laplacian method detailed in Section \ref{sec5.3}.

\indent \textbf{Setup:} To ensure a fair comparison and maintain consistency throughout our study, the experimental configurations for LRNOFF-Fast, including the data splitting ratio, optimizer settings (learning rate and weight decay), and early stopping criteria, are kept identical to those specified for the NOFF in Section \ref{sec5.3}. For the low-rank filter dictionary, to prevent the decomposed sub-filters from converging to identical states while preserving the initial all-pass property, we initialize the parameter matrices with microscopic Gaussian noise. Specifically, the base spectral matrix $\mathbf{B} \in \mathbb{R}^{d \times N}$ and the weight matrix $\mathbf{W} \in \mathbb{R}^{N \times d}$ are initialized with independent and identically distributed (i.i.d.) samples as $B_{k,j} \sim \mathcal{N}(1.0, 0.1^2)$ and $W_{i,k} \sim \mathcal{N}(1/d, 0.01^2)$, respectively.

\indent To demonstrate the necessity and superiority of the low-rank approximation, Table \ref{tab:global_vs_node_large} compares the GFF mechanism and our proposed LRNOFF-Fast across various fractional domains, including GFRFT, MPGFRFT-I, MPGFRFT-II, Lap-GLCT, and wAdj-GLCT. As shown in Table \ref{tab:global_vs_node_large}, the LRNOFF-Fast consistently and substantially outperforms the GFF counterparts. This observation not only verifies that our low-rank decomposition successfully breaks the curse of dimensionality, but also provides strong empirical evidence that node-oriented adaptation is fundamentally superior to traditional global filtering. By assigning customized spectral responses to individual vertices, the node-oriented mechanism effectively overcomes the rigid, one-size-fits-all limitation of global filters. 

\begin{table*}[htbp]
	\centering
	\caption{Denoising performance comparison of GFF-based and LRNOFF-Fast-based methods under various low-rank dimensions $d$ on large-scale datasets.}
	\label{tab:global_vs_node_large}
	\resizebox{\textwidth}{!}{%
		\begin{tabular}{lccc|ccc|ccc|ccc}
			\toprule
			Dataset & \multicolumn{3}{c}{METR} & \multicolumn{3}{c}{PEMS08} & \multicolumn{3}{c}{Solar} & \multicolumn{3}{c}{Electricity} \\
			\cmidrule(lr){2-4} \cmidrule(lr){5-7} \cmidrule(lr){8-10} \cmidrule(lr){11-13}
			$\sigma$ & $40$ & $60$ & $80$ & $100$ & $150$ & $200$ & $8$ & $10$ & $12$ & $3000$ & $5000$ & $6000$ \\
			SNR & $3.495$ & $-0.027$ & $-2.526$ & $8.006$ & $4.485$ & $1.986$ & $1.486$ & $-0.452$ & $-2.036$ & $4.089$ & $-0.348$ & $-1.932$ \\
			\midrule
			& \multicolumn{12}{c}{$d=3$} \\
			\midrule
			GFF-GFRFT & $14.732$ & $13.998$ & $13.439$ & $15.374$ & $14.997$ & $14.478$ & $12.210$ & $11.677$ & $11.249$ & $22.306$ & $20.259$ & $19.106$ \\
			LRNOFF-Fast-GFRFT & $\mathbf{15.960}$ & $\mathbf{14.628}$ & $\mathbf{13.862}$ & $\mathbf{17.931}$ & $\mathbf{15.688}$ & $\mathbf{15.438}$ & $\mathbf{13.306}$ & $\mathbf{12.843}$ & $\mathbf{12.201}$ & $\mathbf{24.256}$ & $\mathbf{21.465}$ & $\mathbf{20.510}$ \\ \midrule
			GFF-MPGFRFT-I & $15.257$ & $14.236$ & $13.426$ & $16.724$ & $15.566$ & $14.844$ & $12.148$ & $11.418$ & $10.882$ & $23.580$ & $20.755$ & $19.906$ \\
			LRNOFF-Fast-MPGFRFT-I & $\mathbf{15.790}$ & $\mathbf{14.565}$ & $\mathbf{13.792}$ & $\mathbf{18.165}$ & $\mathbf{15.964}$ & $\mathbf{15.614}$ & $\mathbf{13.311}$ & $\mathbf{12.633}$ & $\mathbf{12.171}$ & $\mathbf{24.352}$ & $\mathbf{21.544}$ & $\mathbf{20.534}$ \\
			GFF-MPGFRFT-II & $14.819$ & $13.796$ & $13.319$ & $16.326$ & $15.104$ & $14.438$ & $11.934$ & $11.504$ & $11.365$ & $23.733$ & $21.030$ & $20.097$ \\
			LRNOFF-Fast-MPGFRFT-II & $\mathbf{15.803}$ & $\mathbf{14.433}$ & $\mathbf{13.749}$ & $\mathbf{18.208}$ & $\mathbf{16.696}$ & $\mathbf{15.654}$ & $\mathbf{13.577}$ & $\mathbf{12.714}$ & $\mathbf{11.763}$ & $\mathbf{24.242}$ & $\mathbf{21.543}$ & $\mathbf{20.521}$ \\ \midrule
			GFF-Lap-GLCT & $14.728$ & $14.001$ & $13.416$ & $14.897$ & $13.916$ & $14.484$ & $12.987$ & $11.685$ & $11.319$ & $18.252$ & $17.841$ & $19.109$ \\
			LRNOFF-Fast-Lap-GLCT & $\mathbf{15.797}$ & $\mathbf{14.647}$ & $\mathbf{13.844}$ & $\mathbf{17.955}$ & $\mathbf{16.387}$ & $\mathbf{15.443}$ & $\mathbf{13.323}$ & $\mathbf{12.580}$ & $\mathbf{12.216}$ & $\mathbf{24.200}$ & $\mathbf{21.413}$ & $\mathbf{20.471}$ \\
			GFF-wAdj-GLCT & $14.406$ & $13.594$ & $13.007$ & $14.688$ & $13.832$ & $13.178$ & $11.075$ & $10.284$ & $9.738$ & $23.151$ & $20.467$ & $19.501$ \\
			LRNOFF-Fast-wAdj-GLCT & $\mathbf{15.763}$ & $\mathbf{14.460}$ & $\mathbf{13.604}$ & $\mathbf{17.864}$ & $\mathbf{16.334}$ & $\mathbf{15.671}$ & $\mathbf{13.348}$ & $\mathbf{12.534}$ & $\mathbf{12.033}$ & $\mathbf{23.763}$ & $\mathbf{21.239}$ & $\mathbf{20.296}$ \\
			\midrule
			& \multicolumn{12}{c}{$d=5$} \\
			\midrule
			GFF-GFRFT & $14.732$ & $13.998$ & $13.439$ & $15.374$ & $14.997$ & $14.478$ & $12.210$ & $11.677$ & $11.249$ & $22.306$ & $20.259$ & $19.106$ \\
			LRNOFF-Fast-GFRFT & $\mathbf{16.115}$ & $\mathbf{14.656}$ & $\mathbf{13.794}$ & $\mathbf{18.258}$ & $\mathbf{16.063}$ & $\mathbf{15.930}$ & $\mathbf{13.520}$ & $\mathbf{12.738}$ & $\mathbf{12.090}$ & $\mathbf{24.323}$ & $\mathbf{21.569}$ & $\mathbf{20.554}$ \\ \midrule
			GFF-MPGFRFT-I & $15.257$ & $14.236$ & $13.426$ & $16.724$ & $15.566$ & $14.844$ & $12.148$ & $11.418$ & $10.882$ & $23.580$ & $20.755$ & $19.906$ \\
			LRNOFF-Fast-MPGFRFT-I & $\mathbf{15.907}$ & $\mathbf{14.563}$ & $\mathbf{13.781}$ & $\mathbf{18.430}$ & $\mathbf{16.522}$ & $\mathbf{15.933}$ & $\mathbf{13.639}$ & $\mathbf{12.599}$ & $\mathbf{11.943}$ & $\mathbf{24.416}$ & $\mathbf{21.581}$ & $\mathbf{20.646}$ \\
			GFF-MPGFRFT-II & $14.819$ & $13.796$ & $13.319$ & $16.326$ & $15.104$ & $14.438$ & $11.934$ & $11.504$ & $11.365$ & $23.733$ & $21.030$ & $20.097$ \\
			LRNOFF-Fast-MPGFRFT-II & $\mathbf{15.954}$ & $\mathbf{14.529}$ & $\mathbf{13.787}$ & $\mathbf{18.331}$ & $\mathbf{16.628}$ & $\mathbf{15.877}$ & $\mathbf{13.301}$ & $\mathbf{12.464}$ & $\mathbf{12.009}$ & $\mathbf{24.336}$ & $\mathbf{21.588}$ & $\mathbf{20.462}$ \\ \midrule
			GFF-Lap-GLCT & $14.728$ & $14.001$ & $13.416$ & $14.897$ & $13.916$ & $14.484$ & $12.987$ & $11.685$ & $11.319$ & $18.252$ & $17.841$ & $19.109$ \\
			LRNOFF-Fast-Lap-GLCT & $\mathbf{15.842}$ & $\mathbf{14.594}$ & $\mathbf{13.789}$ & $\mathbf{18.267}$ & $\mathbf{16.399}$ & $\mathbf{15.930}$ & $\mathbf{13.502}$ & $\mathbf{12.694}$ & $\mathbf{12.102}$ & $\mathbf{24.437}$ & $\mathbf{21.545}$ & $\mathbf{20.567}$ \\
			GFF-wAdj-GLCT & $14.406$ & $13.594$ & $13.007$ & $14.688$ & $13.832$ & $13.178$ & $11.075$ & $10.284$ & $9.738$ & $23.151$ & $20.467$ & $19.501$ \\
			LRNOFF-Fast-wAdj-GLCT & $\mathbf{15.764}$ & $\mathbf{14.415}$ & $\mathbf{13.596}$ & $\mathbf{18.125}$ & $\mathbf{16.062}$ & $\mathbf{15.697}$ & $\mathbf{13.552}$ & $\mathbf{12.754}$ & $\mathbf{12.131}$ & $\mathbf{24.030}$ & $\mathbf{21.219}$ & $\mathbf{20.311}$ \\
			\midrule
			& \multicolumn{12}{c}{$d=10$} \\
			\midrule
			GFF-GFRFT & $14.732$ & $13.998$ & $13.439$ & $15.374$ & $14.997$ & $14.478$ & $12.210$ & $11.677$ & $11.249$ & $22.306$ & $20.259$ & $19.106$ \\
			LRNOFF-Fast-GFRFT & $\mathbf{16.070}$ & $\mathbf{14.620}$ & $\mathbf{13.810}$ & $\mathbf{18.493}$ & $\mathbf{16.119}$ & $\mathbf{15.914}$ & $\mathbf{13.365}$ & $\mathbf{12.665}$ & $\mathbf{12.112}$ & $\mathbf{24.329}$ & $\mathbf{21.387}$ & $\mathbf{20.374}$ \\ \midrule
			GFF-MPGFRFT-I & $15.257$ & $14.236$ & $13.426$ & $16.724$ & $15.566$ & $14.844$ & $12.148$ & $11.418$ & $10.882$ & $23.580$ & $20.755$ & $19.906$ \\
			LRNOFF-Fast-MPGFRFT-I & $\mathbf{15.906}$ & $\mathbf{14.575}$ & $\mathbf{13.796}$ & $\mathbf{18.502}$ & $\mathbf{16.553}$ & $\mathbf{15.893}$ & $\mathbf{13.241}$ & $\mathbf{12.494}$ & $\mathbf{11.944}$ & $\mathbf{24.313}$ & $\mathbf{21.624}$ & $\mathbf{20.622}$ \\
			GFF-MPGFRFT-II & $14.819$ & $13.796$ & $13.319$ & $16.326$ & $15.104$ & $14.438$ & $11.934$ & $11.504$ & $11.365$ & $23.733$ & $21.030$ & $20.097$ \\
			LRNOFF-Fast-MPGFRFT-II & $\mathbf{15.816}$ & $\mathbf{14.434}$ & $\mathbf{13.655}$ & $\mathbf{18.395}$ & $\mathbf{16.520}$ & $\mathbf{15.926}$ & $\mathbf{13.613}$ & $\mathbf{12.593}$ & $\mathbf{11.708}$ & $\mathbf{24.292}$ & $\mathbf{21.609}$ & $\mathbf{20.518}$ \\ \midrule
			GFF-Lap-GLCT & $14.728$ & $14.001$ & $13.416$ & $14.897$ & $13.916$ & $14.484$ & $12.987$ & $11.685$ & $11.319$ & $18.252$ & $17.841$ & $19.109$ \\
			LRNOFF-Fast-Lap-GLCT & $\mathbf{16.113}$ & $\mathbf{14.647}$ & $\mathbf{13.806}$ & $\mathbf{18.427}$ & $\mathbf{16.412}$ & $\mathbf{15.882}$ & $\mathbf{13.600}$ & $\mathbf{12.675}$ & $\mathbf{12.106}$ & $\mathbf{24.233}$ & $\mathbf{21.472}$ & $\mathbf{20.456}$ \\
			GFF-wAdj-GLCT & $14.406$ & $13.594$ & $13.007$ & $14.688$ & $13.832$ & $13.178$ & $11.075$ & $10.284$ & $9.738$ & $23.151$ & $20.467$ & $19.501$ \\
			LRNOFF-Fast-wAdj-GLCT & $\mathbf{15.812}$ & $\mathbf{14.423}$ & $\mathbf{13.616}$ & $\mathbf{18.358}$ & $\mathbf{16.247}$ & $\mathbf{15.838}$ & $\mathbf{13.586}$ & $\mathbf{12.698}$ & $\mathbf{12.066}$ & $\mathbf{24.245}$ & $\mathbf{21.265}$ & $\mathbf{20.262}$ \\
			\midrule
			& \multicolumn{12}{c}{$d=15$} \\
			\midrule
			GFF-GFRFT & $14.732$ & $13.998$ & $13.439$ & $15.374$ & $14.997$ & $14.478$ & $12.210$ & $11.677$ & $11.249$ & $22.306$ & $20.259$ & $19.106$ \\
			LRNOFF-Fast-GFRFT & $\mathbf{15.945}$ & $\mathbf{14.553}$ & $\mathbf{13.788}$ & $\mathbf{18.577}$ & $\mathbf{16.095}$ & $\mathbf{15.950}$ & $\mathbf{13.319}$ & $\mathbf{12.520}$ & $\mathbf{12.091}$ & $\mathbf{24.220}$ & $\mathbf{21.341}$ & $\mathbf{20.402}$ \\ \midrule
			GFF-MPGFRFT-I & $15.257$ & $14.236$ & $13.426$ & $16.724$ & $15.566$ & $14.844$ & $12.148$ & $11.418$ & $10.882$ & $23.580$ & $20.755$ & $19.906$ \\
			LRNOFF-Fast-MPGFRFT-I & $\mathbf{15.944}$ & $\mathbf{14.512}$ & $\mathbf{13.780}$ & $\mathbf{18.559}$ & $\mathbf{16.535}$ & $\mathbf{15.987}$ & $\mathbf{13.423}$ & $\mathbf{12.663}$ & $\mathbf{12.122}$ & $\mathbf{24.217}$ & $\mathbf{21.493}$ & $\mathbf{20.460}$ \\
			GFF-MPGFRFT-II & $14.819$ & $13.796$ & $13.319$ & $16.326$ & $15.104$ & $14.438$ & $11.934$ & $11.504$ & $11.365$ & $23.733$ & $21.030$ & $20.097$ \\
			LRNOFF-Fast-MPGFRFT-II & $\mathbf{15.865}$ & $\mathbf{14.380}$ & $\mathbf{13.713}$ & $\mathbf{18.524}$ & $\mathbf{16.714}$ & $\mathbf{15.723}$ & $\mathbf{13.429}$ & $\mathbf{12.523}$ & $\mathbf{11.889}$ & $\mathbf{24.255}$ & $\mathbf{21.426}$ & $\mathbf{20.390}$ \\ \midrule
			GFF-Lap-GLCT & $14.728$ & $14.001$ & $13.416$ & $14.897$ & $13.916$ & $14.484$ & $12.987$ & $11.685$ & $11.319$ & $18.252$ & $17.841$ & $19.109$ \\
			LRNOFF-Fast-Lap-GLCT & $\mathbf{16.018}$ & $\mathbf{14.517}$ & $\mathbf{13.800}$ & $\mathbf{18.549}$ & $\mathbf{16.107}$ & $\mathbf{15.961}$ & $\mathbf{13.316}$ & $\mathbf{12.361}$ & $\mathbf{11.955}$ & $\mathbf{24.243}$ & $\mathbf{21.430}$ & $\mathbf{20.372}$ \\
			GFF-wAdj-GLCT & $14.406$ & $13.594$ & $13.007$ & $14.688$ & $13.832$ & $13.178$ & $11.075$ & $10.284$ & $9.738$ & $23.151$ & $20.467$ & $19.501$ \\
			LRNOFF-Fast-wAdj-GLCT & $\mathbf{15.744}$ & $\mathbf{14.435}$ & $\mathbf{13.655}$ & $\mathbf{18.340}$ & $\mathbf{15.981}$ & $\mathbf{15.809}$ & $\mathbf{13.563}$ & $\mathbf{12.705}$ & $\mathbf{12.066}$ & $\mathbf{24.224}$ & $\mathbf{21.346}$ & $\mathbf{20.401}$ \\
			\bottomrule
		\end{tabular}%
	}
\end{table*}

\indent Following the quantitative results in Table \ref{tab:global_vs_node_large}, we provide visualizations to demonstrate how the LRNOFF-Fast mechanism efficiently distills complex spectral responses into a compact yet expressive subspace.

\begin{figure}[htbp]
	\centering
	\includegraphics[width=0.45\textwidth]{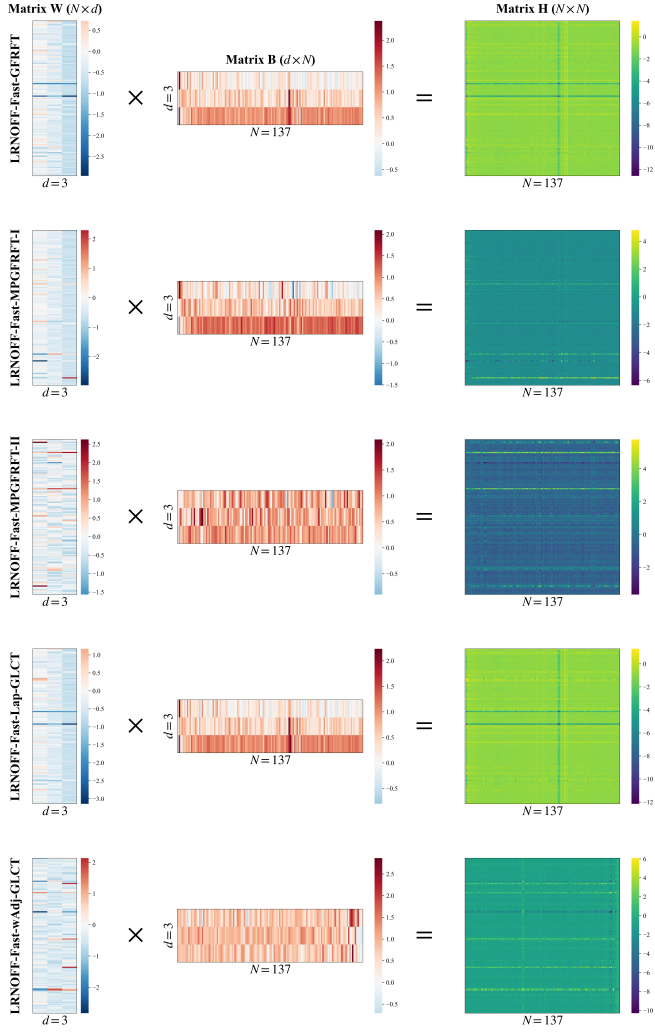}
	\caption{Visualizations of the learned low-rank components on the Solar dataset across various fractional domains.}
	\label{fig:LFNOFF filters}
\end{figure}

\indent 1) Effectiveness of low-rank factorization: First, we visualize the learned decomposition components ($\mathbf{W}$ and $\mathbf{B}$) alongside the reconstructed full filter ($\mathbf{H} = \mathbf{WB}$) on the Solar dataset with a rank of $d=3$. As illustrated in Fig. \ref{fig:LFNOFF filters}, learning $\mathbf{W}$ and $\mathbf{B}$ circumvents the prohibitive $\mathcal{O}(N^2)$ parameter explosion associated with the full filter $\mathbf{H}$. Despite the extreme dimensional compression ($d \ll N$), the reconstructed $\mathbf{H}$ continues to exhibit highly intricate, localized, and domain-specific structural patterns across different fractional transforms. This proves that the low-rank factorization effectively resolves the curse of dimensionality without sacrificing the rich spatial-spectral interactions required for large-scale graphs.

\begin{figure}[htbp]
	\centering
	\includegraphics[width=0.45\textwidth]{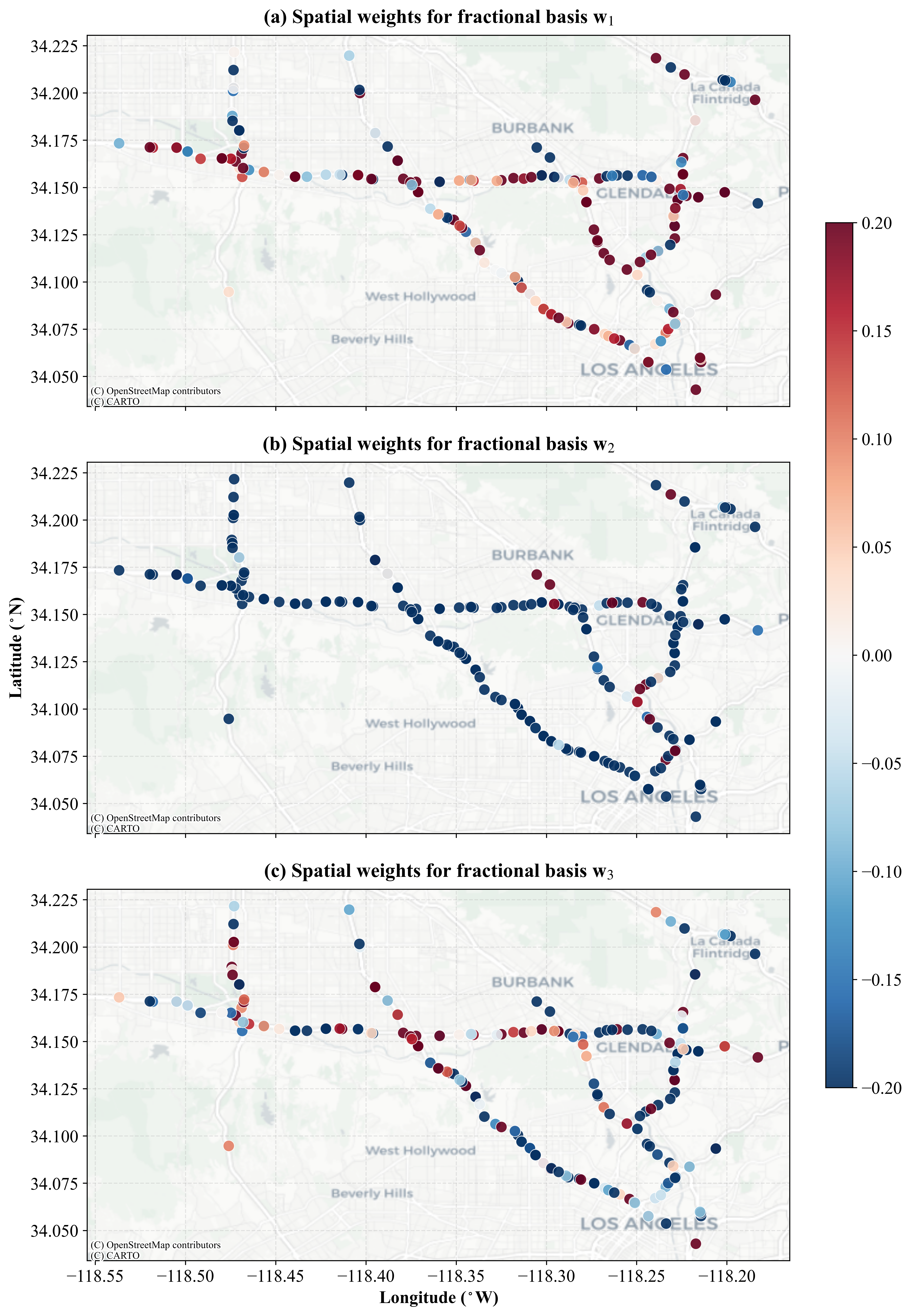}
	\caption{Geographic mapping of the learned spatial weights for the three fractional bases ($\mathbf{w}_1, \mathbf{w}_2, \mathbf{w}_3$) on the METR road network. }
	\label{fig:W}
\end{figure}

\indent 2) Spatial interpretability of the latent subspace: Second, we map the learned spatial weight matrix $\mathbf{W}$ onto the physical geographic space. Using the METR dataset, Fig. \ref{fig:W} illustrates the spatial activation amplitudes of the three latent basis components ($\mathbf{w}_1, \mathbf{w}_2, \mathbf{w}_3$). The visualization reveals a clear separation of spatial functional modes: $\mathbf{w}_2$ exhibits a relatively uniform distribution capturing the global low-frequency trend; $\mathbf{w}_1$ displays distinct regional clustering extracting regional structural features; and $\mathbf{w}_3$ presents rapid spatial variations around complex intersections, acting as a local bottleneck mode. This indicates that the LRNOFF framework successfully learns physically meaningful spatial receptive fields purely driven by data.

\begin{figure}[htbp]
	\centering
	\subfloat[]{
		\includegraphics[width=0.48\linewidth]{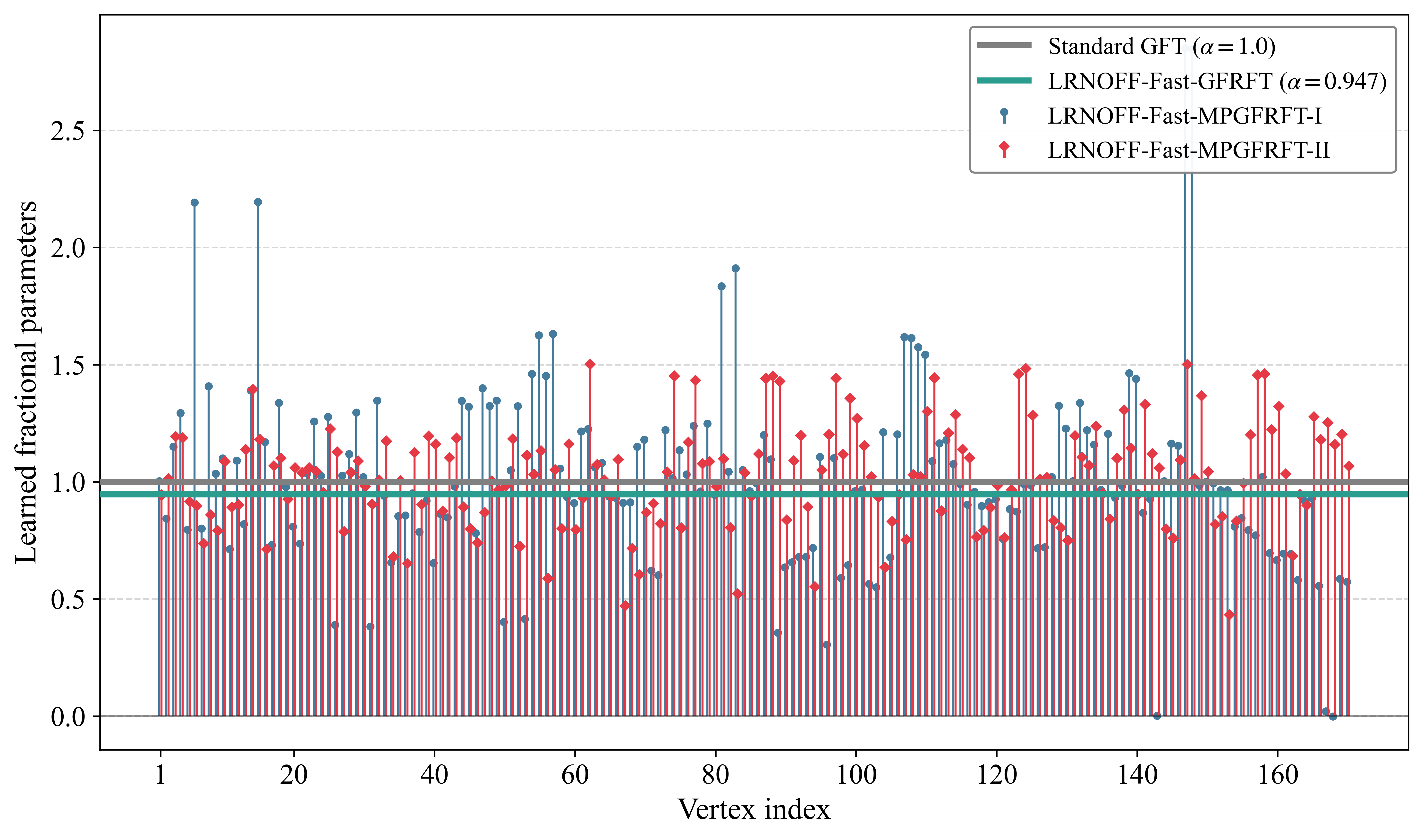}%
		\label{fig:large_param_a}%
	}%
	\hfill 
	\subfloat[]{
		\includegraphics[width=0.48\linewidth]{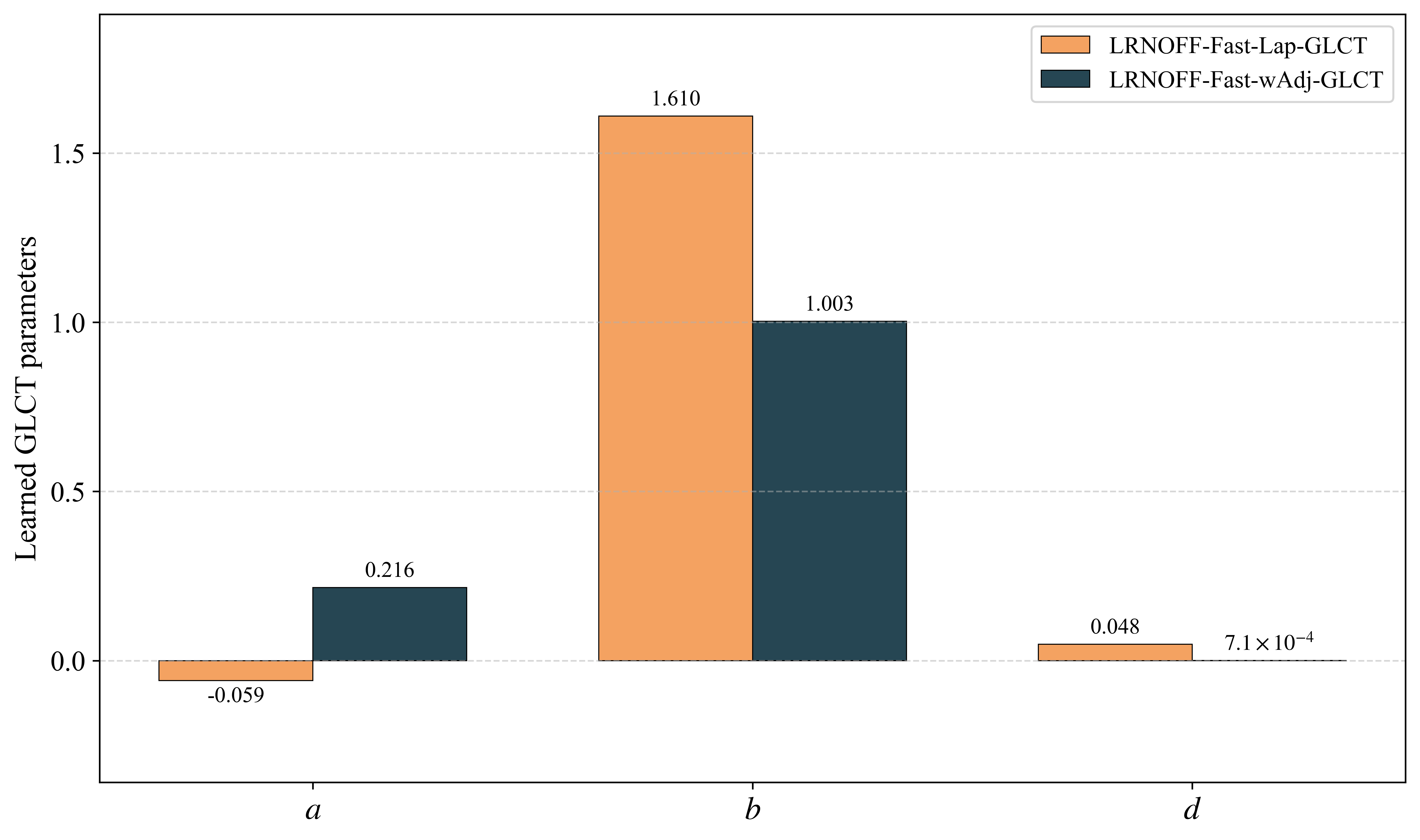}%
		\label{fig:large_param_b}%
	}
	
	\caption{Distribution of the optimized fractional parameters on the PEMS08 dataset ($\sigma=150$, Layer 2). (a) Learned parameters for GFRFT and MPGFRFT variants. (b) Learned phase parameters ($a, b, d$) for GLCT variants. }
	\label{fig:large_learned_parameters}
\end{figure}

\indent 3) Data-driven adaptation of the fractional spectral domains: Finally, to demonstrate how our framework adaptively optimizes the spectral representation, we visualize the learned fractional parameters across five specific transform variants (Fig. \ref{fig:large_learned_parameters}). The results reveal that the optimized parameter sets diverge significantly from the standard Fourier domain ($\alpha=1$). This substantial deviation confirms that our framework effectively transforms the graph spectrum into an optimal fractional domain, maximizing the separation of complex signal and noise representations prior to applying the node-specific filters constructed from a shared set of latent bases.

%
%

\indent Furthermore, to comprehensively evaluate the superiority of our framework on large-scale graphs, we extract the best-performing LRNOFF-Fast configurations from the fractional domains specifically under the low-rank subspace dimension of $d=15$. These optimal variants are then compared against the LRNO-GFT as well as several state-of-the-art neural network baselines. As reported in Table \ref{tab:fractional_vs_others_large}, with the minor exception of a performance tie between LRNOFF-Fast-GFRFT and LRNO-GFT on the Electricity dataset ($\sigma=3000$), the proposed fractional methods exhibit a clear and consistent advantage over the standard GFT approach across all other evaluated scenarios. As previously observed in the small-graph scenarios, the standard GFT suffers from spectral rigidity. This limitation is amplified on massive graphs, where signal components and complex noise patterns are more densely packed. The fractional domains successfully resolve this bottleneck by utilizing their dynamically learned parameters to map the entangled graph signals into an optimal transform space for cleaner separation. Most importantly, our LRNOFF-Fast achieves stronger denoising performance than the neural baselines, perfectly highlighting its remarkable efficiency, anti-overfitting capability, and mathematical interpretability in processing large-scale graph signals.

\begin{table*}[htbp]
	\centering
	\caption{Denoising performance comparison of LRNOFF-Fast ($d=15$) and state-of-the-art baselines on large-scale datasets.}
	\label{tab:fractional_vs_others_large}
	\resizebox{\textwidth}{!}{%
		\begin{tabular}{lccc|ccc|ccc|ccc}
			\toprule
			Dataset & \multicolumn{3}{c}{METR} & \multicolumn{3}{c}{PEMS08} & \multicolumn{3}{c}{Solar} & \multicolumn{3}{c}{Electricity} \\
			\cmidrule(lr){2-4} \cmidrule(lr){5-7} \cmidrule(lr){8-10} \cmidrule(lr){11-13}
			$\sigma$ & $40$ & $60$ & $80$ & $100$ & $150$ & $200$ & $8$ & $10$ & $12$ & $3000$ & $5000$ & $6000$ \\
			SNR & $3.495$ & $-0.027$ & $-2.526$ & $8.006$ & $4.485$ & $1.986$ & $1.486$ & $-0.452$ & $-2.036$ & $4.089$ & $-0.348$ & $-1.932$ \\
			\midrule
			APPNP & $12.419$ & $11.974$ & $11.831$ & $10.037$ & $7.968$ & $6.920$ & $5.705$ & $4.493$ & $3.665$ & $13.855$ & $11.180$ & $5.097$ \\
			GAT & $12.626$ & $12.139$ & $11.984$ & $9.939$ & $7.965$ & $6.945$ & $5.768$ & $4.570$ & $3.752$ & $10.683$ & $9.272$ & $8.252$ \\
			GCN & $12.402$ & $11.947$ & $11.807$ & $9.865$ & $7.893$ & $6.889$ & $5.673$ & $4.472$ & $3.657$ & $10.154$ & $8.850$ & $8.305$ \\
			H2GCN & $12.329$ & $11.547$ & $11.144$ & $10.622$ & $8.805$ & $7.780$ & $8.303$ & $7.394$ & $6.697$ & $9.008$ & $5.577$ & $4.515$ \\
			Chebynet & $12.577$ & $12.120$ & $11.959$ & $10.131$ & $8.065$ & $7.015$ & $5.835$ & $4.629$ & $3.804$ & $14.118$ & $11.323$ & $10.177$ \\
			Bern & $6.080$ & $5.001$ & $4.658$ & $9.407$ & $6.955$ & $5.567$ & $5.349$ & $4.198$ & $3.415$ & $7.955$ & $4.745$ & $3.865$ \\
			GPR-GNN & $13.033$ & $12.138$ & $11.787$ & $10.838$ & $9.047$ & $8.105$ & $8.970$ & $8.025$ & $7.301$ & $14.277$ & $11.423$ & $10.456$ \\
			ARMAcov & $12.615$ & $12.110$ & $11.983$ & $10.103$ & $8.040$ & $6.998$ & $5.828$ & $4.613$ & $3.792$ & $13.902$ & $11.360$ & $10.245$ \\
			Specformer & $12.575$ & $12.205$ & $12.143$ & $11.176$ & $9.632$ & $8.853$ & $5.370$ & $4.481$ & $3.671$ & $7.837$ & $5.626$ & $5.055$ \\
			Unimp & $12.615$ & $12.110$ & $11.902$ & $10.089$ & $8.028$ & $6.986$ & $5.816$ & $4.601$ & $3.790$ & $14.044$ & $11.256$ & $3.875$ \\
			Polyformer & $13.221$ & $12.470$ & $12.207$ & $7.225$ & $6.344$ & $5.954$ & $9.662$ & $8.631$ & $7.500$ & $13.841$ & $11.562$ & $10.564$ \\
			\midrule
			LRNO-Fast-GFT & $15.918$ & $14.416$ & $13.647$ & $18.545$ & $15.765$ & $15.907$ & $13.179$ & $12.209$ & $11.642$ & $24.220$ & $21.315$ & $20.394$ \\
			LRNOFF-Fast-GFRFT & $15.945$ & $\mathbf{14.553}$ & $13.788$ & $\mathbf{18.577}$ & $16.095$ & $15.950$ & $12.210$ & $11.677$ & $11.249$ & $24.220$ & $21.341$ & $20.402$ \\
			LRNOFF-Fast-MPGFRFT & $15.944$ & $14.512$ & $13.780$ & $18.559$ & $\mathbf{16.714}$ & $\mathbf{15.987}$ & $13.429$ & $12.663$ & $\mathbf{12.122}$ & $\mathbf{24.255}$ & $\mathbf{21.493}$ & $\mathbf{20.460}$ \\
			LRNOFF-Fast-GLCT & $\mathbf{16.018}$ & $14.517$ & $\mathbf{13.800}$ & $18.549$ & $16.107$ & $15.961$ & $\mathbf{13.563}$ & $\mathbf{12.705}$ & $12.066$ & $24.243$ & $21.430$ & $20.401$ \\
			\bottomrule
		\end{tabular}
	}
\end{table*}

\section{Conclusion} \label{sec6} 
\indent In this paper, we addressed the fundamental dilemma between spatial adaptability and spectral flexibility in graph signal denoising by proposing a generalized NOFF framework. By introducing node-wise spectral modulation into fractional-domain transforms, NOFF effectively overcomes both the spectral rigidity of the standard node-oriented method and the spatial limitations of globally shared fractional filters. To resolve the catastrophic parameter explosion and noise memorization issues inherent in unconstrained local filtering, we introduced the LRNOFF architecture. Our analysis reveals that imposing a strict low-rank constraint on the ideal localized filter matrix acts as a powerful implicit regularizer, effectively forcing the network to learn robust, physically meaningful spectral bases instead of fitting high-frequency random noise. Moreover, the development of LRNOFF-Fast successfully translates this theoretical optimality into an efficient computational approach for large-scale real-world networks. Empirical results validate the superiority of our framework over existing graph filters and advanced graph neural networks. 


\appendices
\section{Proof of Theorem 1} \label{Appendix A}
\indent Substituting the low-rank constraint \eqref{eq:low_rank_H} into \eqref{filter process}, we have
\begin{align}
	\tilde{x}_i = \sum_{j=1}^{N} \mathcal{F}_{i,j}^{-1} \left( \sum_{k=1}^{d} W_{i,k} B_{k,j} \right) (\mathcal{F}\mathbf{x})_{j}.
\end{align}
\indent Utilizing the commutative property of linear summation, we swap the order of operations
\begin{align} \label{eq:swap_sum}
	\tilde{x}_i = \sum_{k=1}^{d} W_{i,k} \left( \sum_{j=1}^{N} \mathcal{F}_{i,j}^{-1} B_{k,j} (\mathcal{F}\mathbf{x})_{j} \right).
\end{align}
\indent Let the inner summation be denoted as the $i$-th element of an intermediate vector $\mathbf{u}_k \in \mathbb{C}^N$. This inner term represents the $k$-th base filtered signal mapped back to the vertex domain:
\begin{align}
	U_{i,k} = (\mathbf{u}_k)_i &= \sum_{j=1}^{N} \mathcal{F}_{i,j}^{-1} \left[ B_{k,j} (\mathcal{F}\mathbf{x})_{j} \right] \nonumber \\
	&= \left( \mathcal{F}^{-1} (\mathbf{b}_k \odot \mathcal{F}\mathbf{x}) \right)_i.
\end{align}
\indent Subsequently, the outer summation in \eqref{eq:swap_sum} degrades into a localized spatial weighting of the $d$ base signals in the vertex domain
\begin{align}
	\tilde{x}_i = \sum_{k=1}^{d} W_{i,k} U_{i,k}.
\end{align}
\indent By vectorizing this operation across all vertices $i$, the global filtered signal is exactly expressed as $\tilde{\mathbf{x}} = \sum_{k=1}^{d} \mathbf{w}_k \odot \mathbf{u}_k$, which completes the proof. $\hfill\blacksquare$
	
\section{Proof of Theorem 2} \label{Appendix B}	
\indent The filtering output at vertex $i$ can be expressed as the inner product 
\begin{align}
	\tilde{x}_i = (\mathbf{w}^{(i)})^{\top} \mathbf{u}^{(i)}.
\end{align}

\indent Expanding the MSE cost function $J(\mathbf{w}^{(i)}) = \mathbb{E}[|\tilde{x}_i - x_i|^2]$ and setting the partial derivative with respect to $\mathbf{w}^{(i)}$ to zero yields
\begin{align}
	\mathbb{E}\left[(\mathbf{u}^{(i)})^*(\mathbf{u}^{(i)})^{\top}\right] \mathbf{w}^{(i)} = \mathbb{E}\left[(\mathbf{u}^{(i)})^* x_i\right].
\end{align}

\indent Solving this equation directly yields the closed-form Wiener solution, which completes the proof. $\hfill\blacksquare$

\section{Proof of Theorem 3} \label{Appendix C}	
\indent Given the fixed $\mathbf{W}$, the $k$-th base signal is $\mathbf{u}_k = \mathbf{\Psi} \mathbf{b}^{(k)}$. The output at node $i$ can be written as
\begin{align}
	\tilde{x}_i = \sum_{k=1}^{d} W_{i,k} \mathbf{\Psi}_{i,:} \mathbf{b}^{(k)}.
\end{align}
\indent Utilizing the concatenated definition, this simplifies to the inner product $\tilde{x}_i = \mathbf{v}^{(i)} \mathbf{b}$.
Expanding the global MSE cost function $J(\mathbf{b})$ and setting the partial derivative with respect to $\mathbf{b}$ to zero yields
\begin{align}
	\Big( \sum_{i=1}^{N} \mathbb{E}\big[ (\mathbf{v}^{(i)})^{\mathrm{H}} \mathbf{v}^{(i)} \big] \Big) \mathbf{b} = \sum_{i=1}^{N} \mathbb{E}\big[ x_i (\mathbf{v}^{(i)})^{\mathrm{H}} \big].
\end{align}
By substituting the definitions of the global auto-correlation matrix $\mathbf{R}_V$ and the global cross-correlation vector $\mathbf{r}_{Vx}$, the closed-form global Wiener solution is directly obtained. 
$\hfill\blacksquare$

\section{Proof of Theorem 4} \label{Appendix D}	
\indent According to the Eckart-Young-Mirsky theorem, the closest rank-$d$ approximation of any matrix in the Frobenius norm is strictly achieved by truncating its SVD to the principal $d$ components, yielding $\mathbf{H}_{d}^{\star} = \mathbf{U}_d \mathbf{\Sigma}_d \mathbf{V}_d^{\mathrm{H}}$. By introducing the symmetric splitting of the diagonal singular value matrix $\mathbf{\Sigma}_d = \mathbf{\Sigma}_d^{1/2} \mathbf{\Sigma}_d^{1/2}$, the optimal low-rank matrix can be rewritten as $\mathbf{H}_{d}^{\star} = (\mathbf{U}_d \mathbf{\Sigma}_d^{1/2}) (\mathbf{\Sigma}_d^{1/2} \mathbf{V}_d^{\mathrm{H}})$. Substituting the definitions of $\mathbf{W}_{opt}$ and $\mathbf{B}_{opt}$ directly yields $\mathbf{H}_{d}^{\star} = \mathbf{W}_{opt} \mathbf{B}_{opt}$, which completes the proof. $\hfill\blacksquare$	
	


	\bibliography{mybib}
	\bibliographystyle{IEEEtran}

\end{document}